\documentclass[journal]{IEEEtran}
\usepackage[pdftex]{graphicx}
\usepackage{cite}
\usepackage{amsmath}
\usepackage{mathtools}
\usepackage{enumerate}
\usepackage{adjustbox}
\usepackage{algorithm}
\usepackage{algpseudocode}
\newcommand{\doi}[1]{\textsc{doi}: \href{http://dx.doi.org/#1}{\nolinkurl{#1}}}
\usepackage{stfloats}
\usepackage{footnote}
\usepackage[inline]{enumitem}
\usepackage{microtype} 
\usepackage{acronym}
\usepackage{xcolor}
\usepackage{colortbl,hhline}
\usepackage{tabularx}
\usepackage{boldline}
\usepackage{multirow,tabu}
\usepackage{tikz,pgfplots}
\usepackage{subcaption}
\usetikzlibrary{arrows}
\usepackage{pgfplotstable}
\usepackage{fancyhdr}

\newcommand{\ourtool}[0]{\textsc{\acs{aads}}}
\newcommand{\ourtoolext}[0]{\textsc{\acs{daics}}}

\newcommand{\keepcomment}{1} 
\ifnum\keepcomment=1
	\newcommand{\roberto}[1]{{\leavevmode\color{blue}{[ROB: #1]}}}
	\newcommand{\domenico}[1]{{\leavevmode\color{magenta}{[DOM: #1]}}}
	\newcommand{\maged}[1]{{\leavevmode\color{orange}{[MAG: #1]}}}
\else
	\newcommand{\roberto}[1]{\ignorespaces\unskip}
	\newcommand{\domenico}[1]{\ignorespaces\unskip}
	\newcommand{\maged}[1]{\ignorespaces\unskip}
\fi

\newlength{\Oldarrayrulewidth}

\pgfplotsset{
	compat = newest,
	tick label style={font=\sffamily\tiny},
	label style={font=\sffamily\scriptsize},
	legend style={font=\sffamily\small\raggedleft, mark options={mark size=1.5pt}},
	legend cell align=left,
	legend image code/.code={
                            \draw[mark repeat=2,mark phase=2]
                            plot coordinates {
                            (0cm,0cm)
                            (0.2cm,0cm)        
                            (0.4cm,0cm)         
                            };%
                            },
	grid style={dotted,gray}
}

\graphicspath{{../}{graphs/}}
\makeatletter
\renewcommand{\fnum@figure}{Figure \thefigure}
\makeatother
\begin{document}
\bstctlcite{IEEEexample:BSTcontrol} 

\title{\acs{daics}: A Deep Learning Solution for Anomaly Detection in Industrial Control Systems}

\author{
	Maged Abdelaty$^\alpha$$^\beta$,
	Roberto Doriguzzi-Corin$^\alpha$,
	Domenico Siracusa$^\alpha$\\
	
	\small{$^\alpha$ICT, Fondazione Bruno Kessler - Italy}\\
	\small{$^\beta$University of Trento - Italy}
	
}

\maketitle              
\thispagestyle{fancy}
\renewcommand{\headrulewidth}{0pt}
\chead{\scriptsize This work has been submitted to the IEEE for possible publication. Copyright may be transferred without notice, after which this version may no longer be accessible.}

\acrodef{ics}[ICS]{Industrial Control System}
\acrodef{icn}[ICN]{Industrial Control Network}
\acrodef{plc}[PLC]{Programmable Logic Controller}
\acrodef{rnn}[RNN]{Recurrent Neural Network}
\acrodef{svm}[SVM]{Support Vector Machines}
\acrodef{gan}[GAN]{Generative Adversarial Network}
\acrodef{lstm-rnn}[LSTM-RNN]{Long Short-Term Recurrent Neural Network}
\acrodef{cnn}[CNN]{Convolutional Neural Network}
\acrodef{mlp}[MLP]{Multilayer Perceptron}
\acrodef{swat}[SWaT]{Secure Water Treatment}
\acrodef{fpr}[FPR]{False Positive Rate}
\acrodef{scada}[SCADA]{Supervisory Control and Data	Acquisition}
\acrodef{hmi}[HMI]{Human-Machine Interface}
\acrodef{dl}[DL]{Deep Learning}
\acrodef{mse}[MSE]{Mean Square Error}
\acrodef{sgd}[SGD]{Stochastic Gradient Descend}
\acrodef{aads}[AADS]{Adaptive Anomaly Detection in industrial control Systems}
\acrodef{aads-ext}[AADS-Ext]{Extension-Adaptive Anomaly Detection in industrial control Systems} 
\acrodef{daics}[DAICS]{Deep learning Anomaly detection in Industrial Control Systems} 
\acrodef{wadi}[WADI]{Water Distribution}
\acrodef{uae}[UAE]{Undercomplete Auto-encoder}
\acrodef{pca}[PCA]{Principle Component Analysis}
\acrodef{batadal}[BATADAL]{BATtle of the Attack Detection ALgorithms}
\acrodef{dfft}[DFFT]{Discrete Fast Fourier Transform}
\acrodef{ads}[ADS]{Anomaly Detection System}
\acrodef{ids}[IDS]{Intrusion Detection System}
\acrodef{wdnn}[WDNN]{Wide and Deep Neural Network}
\acrodef{ttnn}[TTNN]{Threshold Tuning Neural Network}
\begin{abstract}

Deep Learning is emerging as an effective technique to detect sophisticated cyber-attacks targeting \acp{ics}. The conventional approach to detection in literature is to learn the ``normal'' behaviour of the system, to be then able to label noteworthy deviations from it as anomalies. However, during operations, \acp{ics} inevitably and continuously evolve their behaviour, due to e.g., replacement of devices, workflow modifications, or other reasons. As a consequence, the accuracy of the anomaly detection process may be dramatically affected with a considerable amount of false alarms being generated.
This paper presents \ourtoolext, a novel deep learning framework with a modular design to fit in large \acp{ics}. The key component of the framework is a 2-branch neural network that learns the changes in the \ac{ics} behaviour with a small number of data samples and a few gradient updates. This is supported by an automatic tuning mechanism of the detection threshold that takes into account the changes in the prediction error under normal operating conditions. In this regard, no specialised human intervention is needed to update the other parameters of the system. \ourtoolext\ has been evaluated using publicly available datasets and shows an increased detection rate and accuracy compared to state of the art approaches, as well as higher robustness to additive noise.



\end{abstract}


\section{Introduction}\label{sec:introduction}

Nefarious cyber-attacks targeting \acfp{ics} may cause service downtime or material losses in industrial production sites, with potential negative repercussions on the lives of citizens. A notable example is the blackout attack against the Ukrainian power grid perpetrated in late 2015 \cite{infrastructure16}, which exploited the BlackEnergy (BE) malware. In that occasion, the attackers intruded remotely into the computers of three regional power distribution companies in a coordinated manner. They executed a malicious code to alter the firmware of specific control devices and to instruct unscheduled disconnections from servers. The attack affected thousands of users and left them without electricity. Another example is the attack the Iranian nuclear program \cite{karnouskos2011stuxnet} in 2010 by means of the Stuxnet work. The worm infected the code running inside the \acp{plc}, collecting information on the \ac{ics} and damaging the centrifuges inside the plant by repeatedly changing their rotation speed.

The increasing occurrence of such attacks and their complexity have motivated the development of intrusion detection solutions for \acp{ics} based on machine learning techniques. Among the different approaches proposed in the scientific literature, a popular and powerful technique is the one-class classification. At the training stage, solutions based on one-class classification build a model that represents the normal behaviour of the \ac{ics} (the class of ``normality''). At the production stage, the detection system uses that model to verify whether the behaviour of the live system matches the expected normal behaviour. Deviations from the normality, usually determined through thresholds on the classification error, are flagged as anomalies.

These algorithms can be very sensitive to abnormal behaviours, including zero-days anomalies or attacks and faulty devices or sensors, since they may evade the detection process. Moreover, real-world \acp{ics} are dynamic and operate in noisy environments: these factors may hamper the correct functioning of the detection system because they can ``shift'' the normal behaviour.
In such scenarios, the main challenges are to ($i$) timely update the model of normality upon changes in the production workflows, and ($ii$) periodically refresh the detection thresholds. Previous works either do not tackle such challenges \cite{lin2018tabor, kravchik2019efficient} or require highly-specialised human intervention to update the parameters of the model \cite{goh2017anomaly, Shalyga2018}.

This paper tackles the aforementioned challenges by proposing \acs{daics} (\acl{daics}), an anomaly detection solution for \acp{ics} based on the one-class classification paradigm. \ourtoolext\ combines a database-oriented management approach and a deep learning architecture to model the normal behaviour of the \ac{ics}. The proposed neural network relies on \textit{wide and deep} learning and \textit{convolutional layers} to memorise and generalise the characteristics of the \ac{ics}. Wide and deep learning is a technique that has been recently proposed to improve the performance of recommender systems \cite{cheng2016wide}. Convolutional layers, introduced by Yann LeCun in 1990 \cite{lecun}, apply small filters to detect patterns in the input while requiring a limited number of trainable parameters. Convolutional layers, and \aclp{cnn} in general, have become very popular in   many application areas, including network security \cite{potluri2018convolutional, lucid,NGUYEN2020418} and anomaly detection in \acp{ics} \cite{Potluri2020,8802785,8903886}. 
  
As anticipated above, a common issue of one-class classification mechanisms is the high false alarm rate caused by the progressive evolution of the normal behaviour of the \ac{ics} with respect to the initial one, based on which the model was trained. This phenomena, called \textit{domain shift}, can be caused by changes in the industrial workflow, degradation of devices and communication links over the time, installation/removal of devices, updates in the devices' firmware or configuration, or to external noise on the communication channels. \ourtoolext\ implements an automatic threshold tuning algorithm and the so-called \textit{few-time-steps} algorithm, the latter based on the few-shot learning paradigm \cite{finn2017model, nichol2018first}, to quickly update the model with the latest changes of the \ac{ics} normal behaviour.

We evaluate \ourtoolext\ using datasets collected from the \ac{swat} and \ac{wadi} testbeds, which are widely used for security research in \acp{ics} \cite{datasets-url}. The datasets comprise a training set containing only normal records, and a test set with normal and anomalous records. The anomalies in the test sets are real-world attacks targeting the integrity and the availability of the testbeds \cite{adepu2016generalized}. \ourtoolext\ has been compared with state-of-the-art solutions in terms of anomaly detection accuracy and robustness to Gaussian noise. 

The contributions of this work can be summarised as follows:
\begin{itemize}
   \item
   An \ac{ads} that has a modular architecture to fit large \acp{ics}. The key components of the proposed \ac{ads} is a 2-branch neural network, in which a \textit{wide branch} is designed to memorise the existing relations between the input features, while a \textit{deep branch} generalises the model to unknown relations.
    \item
    An automatic threshold tuning technique that dynamically tunes the detection threshold based on the prediction error observed on the live \ac{ics}. 
    \item
    A sensitivity analysis of hyperparameters of the proposed \ac{ads} using two real-world datasets, namely, the \ac{swat} and \ac{wadi}. The analysis provides practical insights to improve the performance under different operating conditions. 
\end{itemize}

\ourtoolext\ extends our previous work called \ourtool\ (Adaptive Anomaly Detection in industrial control Systems) \cite{abdelaty2020aads}. With respect to \ourtool, \ourtoolext\ scales better to larger \acp{ics}, adopts an automated threshold tuning mechanism based on deep learning, and it has been tested on two public datasets: \ac{swat} (like \ourtool) and \ac{wadi}. The comparison presented in this paper shows that \ourtoolext\ improves \ourtool\ in terms of precision (fewer false positives) and robustness to additive noise on the \ac{swat} dataset.

The rest of this paper is organised as follows: Section \ref{sec:Related_Work} reviews state-of-the-art works on anomaly detection in \acp{ics}.  Section \ref{sec:threat_model} defines the threat model used in this research. We describe the \ac{swat} and \ac{wadi} datasets in Section \ref{sec:datasets}.
The problem statement is provided in section \ref{sec:Problem_Statement}. Section \ref{sec:Approach} introduces the proposed anomaly detection framework. 
Experimental setup and the evaluation results are presented in section \ref{sec:Results}. We conclude this paper in section \ref{Conclusion}.
\section{Related work} \label{sec:Related_Work}


This section reviews recent research studies on anomaly detection in \acp{ics}. Particular attention is given to those works focusing on water treatment plants and validated using the \ac{swat} and \ac{wadi} datasets, as their performance is discussed in Section \ref{sec:Results} for state-of-the-art comparison.

In \cite{goh2017anomaly}, Goh et al.  present an approach based on a \ac{rnn} and the Cumulative Sum (CuSum) technique to detect the anomalies. CuSum sets upper and lower control limits for the prediction error in each sensor and actuator. An anomaly is detected when the prediction error is outside such limits. Besides suffering from a high false-positive rate, the main limitation of this approach is that it might require an expert human intervention to tune and update the CuSum limits.

Kravchik et al. \cite{Kravchik2018} propose a solution for detecting anomalies using convolutional and recurrent neural networks. The key aspect of this work is a statistical approach for anomaly detection based a normalised value of the prediction error. The normalisation is computed by using the mean and standard deviation of the prediction error of the benign samples recorded during normal operations of the water treatment plant, as recorded in the \ac{swat} dataset. However, the authors do not explain how such statistical properties are updated in the case of changes in the production environment, quite frequent in real-world environments, as discussed in Section \ref{sec:introduction}. An extension of this work can be found at \cite{kravchik2019efficient}, in which the authors present their preliminary results (the paper has not been peer-reviewed at the time of writing) on \ac{swat}, \ac{wadi} and BATADAL datasets. In addition to the 1D-\ac{cnn} proposed in the first paper, the authors explore the performance of two more models such as \ac{uae} and windowed-\ac{pca}. We compare the performance of \ourtoolext\ and these works in Section \ref{sec:Results}. 

The approach presented in \cite{FARSI2019online} is based on computing the distance between current and previous devices states. It is assumed that the distance should stay below a threshold during normal operations. The proposed solution has been tested only on the \ac{swat} dataset, where it shows similar performance to \ourtoolext\ in terms of F1 score, as discussed in Section \ref{sec:Exp_1}. However, the number of detected attacks is not shown in the paper, hence preventing a full understanding of the effectiveness of the approach.  


The anomaly detection solution proposed in \cite{Shalyga2018} is based on an \ac{mlp} and relies on a threshold applied to a weighted sum of the prediction errors of all sensors and actuators. Low weights are assigned to those devices whose normal behaviours are hard to predict. This strategy, adopted to increase the performance on the \ac{swat} dataset, shows a high rate of false negatives, which leads to a low number of detected attacks, as shown in Section \ref{sec:Results}.

TABOR \cite{lin2018tabor} is an anomaly detection solution validated on the \ac{swat} dataset. TABOR combines two different models, namely: Probabilistic Deterministic Finite Automaton (PDFA), and a Bayesian Network (BN). The final anomaly detection is based on a combination of results from the two models. Also in this case, the authors do not address the problem of updating the model in case of changes in the normal operations of the \ac{ics}. Updating the model here appears significantly cumbersome due to the complex characterisation of the interaction between sensors and actuators needed to build the model.

In general, \acp{ads} rely on a threshold that acts as the maximum allowed prediction error during normal operations, above which a systems state is considered anomalous. The threshold is either selected empirically by an expert \cite{Kravchik2018} or based on the prediction error on the training set \cite{Shalyga2018, FARSI2019online}. In this paper, we propose a technique for automated threshold tuning, which dynamically adjusts the threshold on test time, improving the ability of \ourtoolext\ to adapt to the normal behaviour evolution. In \cite{ali2013automated, park2018multimodal}, Markov chains and support vector regression are employed to dynamically tune thresholds for anomaly detection in network traffic (e.g., port scans, brute force attacks, SQL injections, etc.) and robot-assisted feeding. To the best of our knowledge, \ourtoolext\ is the first anomaly detection solution for water treatment plants (and \acp{ics} in general) to implement an automated threshold tuning mechanism. 

In summary, a common drawback of available solutions is that they are not flexible enough to quickly and efficiently adapt to changes in the production environment. In a water treatment plant, examples of such changes are: increasing the size of a water tank or replacing a motorised valve with another with different operation modes. Instead, our approach is based on a novel algorithm called \textit{few-time-steps}, presented in Section \ref{sec:few-time-steps}, that fine-tunes the neural network according to the changes in the normal behaviour of the \ac{ics}. The proposed algorithm uses a small amount of data to update the weights of the neural network and requires minimal human intervention. This is supported by an automated threshold tuning technique, which adds to \ourtoolext\ ability to cope with the dynamic behaviour of industrial environments.
Moreover, the proposed neural network architecture is designed to be  scalable, so it can model the normal behaviour of large \acp{ics}.

\section{Threat Model} \label{sec:threat_model}

We assume that attacker's capabilities include gaining remote access to the networked control system, to the \ac{scada} workstation, or can physically compromise sensors and actuators within the \ac{ics}. We also assume that the attacker has domain knowledge of the targeted \ac{ics}, e.g., the physical property measured by each sensor and the physical consequences of actuation commands. 
The attacker's goal is to use the above capabilities and knowledge to damage or modify the \ac{ics} operations. This includes: (i) altering the state of actuators through a MITM-like attack, in which the actuator receives commands from the attacker instead of a \ac{plc}, (ii) sending spoofed sensors readings to the \ac{plc} to drive the \ac{plc} to take wrong decisions, and (iii) tampering with the \ac{plc} firmware aimed to take the \ac{ics} out of service or to change the programmed logic (e.g., the Stuxnet worm \cite{karnouskos2011stuxnet}). 



\section{Datasets} \label{sec:datasets}
We evaluate \ourtoolext\ using two popular public real-world datasets, namely \acf{swat} and \acf{wadi}. In this section we provide a short introduction to the main properties of the two dataset, also summarised in Table \ref{tab:dataset-comparison}.

\begin{table}[!ht]
	\centering
	\caption{Properties of the \ac{swat} and \ac{wadi} datasets.}
	\label{tab:dataset-comparison}
	\renewcommand{\arraystretch}{1.1}
	\begin{tabular}{l@{\hskip 1cm}c@{\hskip 3mm}c@{\hskip 3mm}c@{\hskip 3mm}c@{\hskip 3mm}c}
		\hline
		\textbf{Dataset} & \textbf{Records} & \begin{tabular}{@{}c@{}}\textbf{Duration}\\\textbf{(days)}\end{tabular} & \textbf{Attacks} & \textbf{Sensors} & \textbf{Actuators} \\ \hline
		\textbf{\ac{swat}} & 946,722 & 11 & 36 & 25 & 26  \\
		\textbf{\ac{wadi}} & 1,209,610 & 16 & 15 & 69 & 54 \\
		\hline
	\end{tabular}
\end{table}

\subsection{The \ac{swat} dataset}
\label{subsec:swat}
This dataset has been collected from the \ac{swat} testbed, a reduced version of an operational clean water treatment plant \cite{Goh2017swat, datasets-url}. The water treatment process is monitored by a \ac{scada} workstation and is divided into six sub-processes, including raw water supply and storage, ultra-filtration and backwash.  The sub-processes are controlled by a pair of \acsp{plc} that communicate with sensors (water flow indication transmitters, level indicator transmitters, analyser indicator transmitters, such as pH analysers, and ultra-violet modules) and actuators (e.g., pumps, motorised valves). The \acp{plc} collect the readings from the sensors that monitor the status of the physical process, and send actuation commands to the actuators. Based on the \ac{plc}'s internal logic, such commands may be used to either change or keep the current state of an actuator. 

The dataset consists of 946,722 records of sensors and actuators collected during 11 days of operation at a rate of one sample per second. Each record contains 51 attributes, representing 25 sensors readings and 26 actuators states. The dataset is divided into a 7-day portion of normal operations, which has been used as the training set, plus a 4-day portion of normal activity combined with 36 attacks generated by following the attack model in \cite{adepu2016generalized} (the test set). 20\% of the training set is used for validation (validation set). 

The attacks in the test set are of duration ranging between two minutes and nine hours, and involve one or multiple devices at the same time. Attacks include spoofing the readings of sensors or reversing the operation states of actuators. For instance, in the first attack, the state of a motorised valve is switched from closed to open for 15 minutes aiming to cause a tank overflow. In another attack, the value reported by a water level sensor is decreased by 0.5mm/sec for seven minutes, again with the intention of causing a tank overflow. A more detailed description of these attacks is available in \cite{Goh2017swat, adepu2016investigation}.

\subsection{The \ac{wadi} dataset}
\label{subsec:wadi}

The \ac{wadi} testbed is a reduced version of a real-world water distribution network \cite{ahmed2017wadi, datasets-url}. The testbed network consists of three sub-processes, or stages, including a primary grid for water supply, a secondary grid for water distribution to six consumer tanks, and a return-water grid that handles the excess of water from the consumer tanks. Similarly to the \ac{swat} testbed, also \ac{wadi} is supervised by a \ac{scada} workstation. Moreover, each sub-process is controlled by a \ac{plc} communicating with sensors and actuators. Sensors include water flow indication transmitters, level indicator transmitters, analyser indicator transmitters and pressure meters. Like the \ac{swat} testbed, actuators include pumps and motorised valves.

The \ac{wadi} dataset comprises 1,209,610 records, each with 123 attributes divided into 69 sensors readings and 54 actuators states collected during a period of 16 days. Normal operation conditions have been recorded during the first 14 days, split into training (95\%) and validation (5\%) sets. The last two days, with normal activity and 15 attacks, are used as test set. The attacks follow the same model used in the \ac{swat} dataset and described in \cite{adepu2016generalized},  although with shorter duration (between 2 and 30 minutes). For instance, one attack cuts off the water supply to consumer tanks; another one aims at increasing the level of chemicals in the water by turning off a sensor and sending false readings to the \ac{plc} for ten minutes. More details and examples on the \ac{wadi} attacks can be found in \cite{ahmed2017wadi}.
\section{Problem Statement}\label{sec:Problem_Statement}
\input{results_data}

Unsupervised anomaly detection solutions for \acp{ics} are usually built using the so-called \textit{one-class classification} technique. The basic idea is to build a model of the normal behaviour of the industrial process and to consider as anomalous every event that does not fit the model. The main challenge with such approaches is dealing with the \textit{domain shift}, which originates from a gap in the data distribution between the training and unseen data \cite{quionero2009dataset, wang2018deep}. The tangible effect of the domain shift is an increase of false alarms generated by the anomaly detection system due to normal events classified as anomalies.

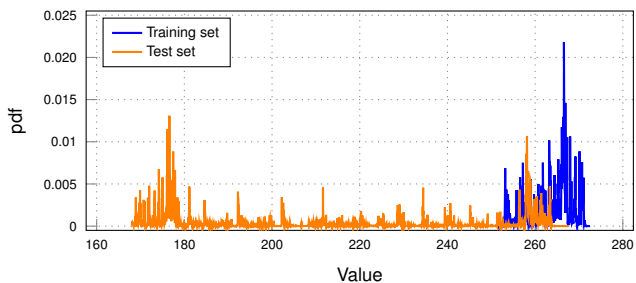
\begin{figure}[h!]
	\centering
	\begin{tikzpicture}
	\begin{axis}[  
	legend style={nodes={scale=0.6}},
	legend pos=north west,
	legend columns=1,
	height=4.5 cm,
	width=1\linewidth,
	grid = both,
	xlabel={Value},
	ylabel={pdf},
	scaled y ticks=false,
	scaled x ticks=false,
	xmin=160,
	xmax=280,
	xtick={160,180,200,220,240,260,280},
	xticklabels={160,180,200,220,240,260,280},
	xtick pos=left,
	ymin=0, ymax=0.025,
	ytick={0,0.005,0.01,0.015,0.020,0.025},
	yticklabels={0,0.005,0.01,0.015,0.020,0.025},
	ytick pos=left,
	enlargelimits=0.02
	]
	\addplot [color=blue,style=thick] table[x index=0,y index=1] {\SensorDensitySevenDays}; 
	\addplot [color=orange,style=thick] table[x index=0,y index=1] {\SensorDensityFourDays}; 
	\legend{Training set, Test set}
	\end{axis}
	\end{tikzpicture}
	\caption{The probability density function for sensor AIT201. It shows the variation in the normal behaviour between training and test sets.}
	\label{fig:AIT201}
\end{figure}

The domain shift problem is present in the \ac{swat} dataset, where we can observe changes in the normal behaviour of some devices across the training and test sets.  For instance, during the normal operation, the pump P102 has a single state of value 1 in the training set, then it takes an additional ``normal'' state of value 2 in the test set. Also, the probability distribution of some sensors changes between the two sets. For example, in the training set the output of the Analyser Indication Transmitter AIT201 ranges in interval [251, 272] $\mu S/cm$ (micro Siemens per centimetre), while in the test set it ranges in [168, 267] $\mu S/cm$ with a substantial different distribution, as illustrated in Figure \ref{fig:AIT201}.
Another non-trivial and representative example is the presence of redundant devices, such as the redundant pump P102 in the \ac{swat} testbed. A redundant pump is always off until the primary pump stops working unexpectedly. In this case, the \ac{plc} turns on the redundant pump instantaneously to take over the work of the primary pump. If such a process is not covered in the training set, the forecasting model will consider the operations of the redundant pump as anomalies.

The domain shift problem can be also observed in the \ac{wadi} dataset. For instance, the records of one of the \textit{Flow Indication Transmitters} range within [0, 2.3] $m^3/hour$ of water in the training set, while the range changes to [0, 3.3] $m^3/hour$ in the test set. 

The characterisation of the normal behavioural evolvement is still a challenge for the implementation of the anomaly detection solutions in real-world systems \cite{mulinka2018adaptive, mulinka2018stream}. Current research in \acp{ics} have tackled the domain shift problem by only focusing on adjusting the parameters of the detection algorithms or excluding devices with domain shift from the modelling process, as explained before in Section \ref{sec:Related_Work}. However, we argue that tuning the detection parameters without updating the model of normality is not sufficient to cope with dynamic environments such as \acp{ics}. 


In the next section, we present our solution for anomaly detection in \acp{ics} that encompasses a lightweight technique, called \textit{few-time-steps learning}, which is inspired by the few-shot learning paradigm \cite{chen2019closer, nichol2018first}. The \textit{few-time-steps learning} technique updates the initial model of the industrial process throughout its evolvement in the hardware and software configuration, hence minimising the number of false positives caused by the domain shift phenomena.
\section{The \ourtoolext\ framework} 
\label{sec:Approach}

\ourtoolext\ defines a model of the normal behaviour of the \ac{ics} combining a neural network with a database-like approach. Deep learning techniques are used to characterise the continuous readings of sensors, whose values are usually sampled by a \ac{plc} very frequently, while the database memorises the states of actuators, as commanded by the \ac{plc}.  \ourtoolext\ includes the few-time-steps learning algorithm to update the model based on the normal behavioural evolvement of the \ac{ics}. The detection of anomalies is determined with a threshold that is automatically adjusted using the prediction error.

\begin{table}[!t]
	\centering
	\scriptsize
	\caption{Glossary of symbols.}
	\renewcommand{\arraystretch}{1.1}
	\label{tab:notations}
	\begin{adjustbox}{width=1\linewidth}
		\begin{tabular}{|l|l|}
			\hline
			\acs{wdnn} & \acl{wdnn}  \\
			\hline
			\acs{ttnn} & \acl{ttnn} \\
			\hline
			$W_{in}$ & Input time window \\
			\hline
			$W_{out}$ & Output time window  \\
			\hline
			$W_{anom}$ & Anomaly waiting time  \\
			\hline
			$H$ & Horizon  \\
			\hline
			$G$ & Number of output section of \acs{wdnn}  \\
			\hline
			$\mu_{out}[g]$ & Output section $g$ of \acs{wdnn} \\
			\hline
			$m_{ac}$ & Total number of actuators\\
			\hline
			$m_{se}$ & Total number of sensors\\
			\hline
			$m_{se}^{g}$ & Number of sensors of output section $g$ \\
			\hline
			$\acs{mse}_{g,t}$ & Prediction error on section $g$ at time $t$  \\
			\hline
			$\acs{mse}_{g,[a,b)}$ & List of prediction errors on section $g$ between time $a$ and $b-1$ \\
			\hline
			$T_g$ & Anomaly threshold for section $g$  \\
			\hline
			$\nu_{t}$ & Actuator states recorded at time $t$ \\
			\hline
		\end{tabular}
	\end{adjustbox}
\end{table}

\subsection{Prediction of sensor states}
\label{subsec:anom_sensors}

\begin{figure*}[t!]
	\centering
	\includegraphics[width=\textwidth]{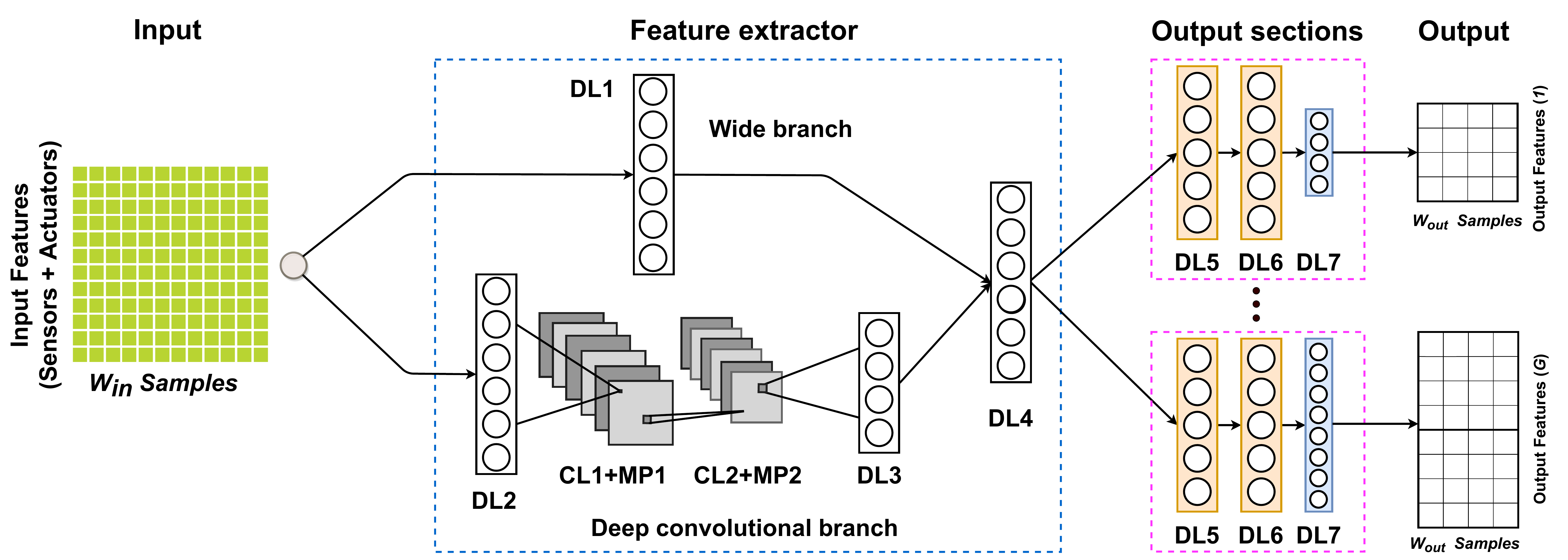}
	\caption{Architecture of \ac{wdnn}. The prediction of the sensors states,i.e. the output features, is computed combining the information from both sensors and actuators. Note that the size of DL7 varies according to the number of output features in each output section.}
	\label{fig:arch_wide_deep_cnn}
\end{figure*}

The normal behaviour of sensors is modelled using a deep learning architecture called \acf{wdnn}. The architecture of \ac{wdnn}, shown in Fig. \ref{fig:arch_wide_deep_cnn}, is inspired by the recommender system introduced in \cite{cheng2016wide}, where the authors trained a wide and deep neural network to recommend apps based on the user's query and preferences. Our neural network has two goals: memorization through the wide branch and generalization through the deep branch. Memorization means learning the relationship between feature-pairs in the training set, hence recording the co-occurrence of combinations of sensors values.  Generalization means the ability to explore relationships that do not exist in the training set.

Like in our previous work \cite{abdelaty2020aads}, the neural network comprises a \textit{Feature extractor} section that learns the relations between all sensors and actuators during the normal operation of an \ac{ics} and extracts the features required to predict the normal states of sensors. Instead, the output section is split into multiple branches to serve large-scale \acp{ics} where the sensors are controlled by several \acp{plc}, usually specialised on a specific part of the industrial process with specific behaviour and anomaly threshold.


\textbf{Architecture.} The neural network takes as input an array $X$ of data samples collected during a time window of length $W_{in}$ seconds, corresponding to $W_{in}$ samples, as the dataset was collected at a sampling rate of one sample per second. The size of $X$ is $m\times W_{in}$, where $m=m_{se}+m_{ac}$ is the number of features for each sample including the state of $m_{se}$ sensors and $m_{ac}$ actuators taken at time $t$. Please note that we use the features from both sensors and actuators because the behaviour of the sensors depends on their current states and the actions taken by actuators. The output $Y$ is the predicted readings from sensors during a future time window $W_{out}$. The two time-windows $W_{in}$ and $W_{out}$ are separated by a time interval called horizon $H$. This separation prevents the forecasting model from copying the last values of the input time window $W_{in}$ into $W_{out}$, as pointed out by the authors of \cite{Shalyga2018}.

As shown in Fig. \ref{fig:arch_wide_deep_cnn}, the neural network architecture comprises two convolutional layers and seven fully connected layers. The fully connected layer DL1 defines the so-called \textit{wide branch} used for the memorisation of the normal state of sensors and actuators using cross products between the input features.


The \textit{deep branch} provides the level of generalisation necessary for correctly handling the events not covered in the training set, aiming to minimise the prediction error in the case of new input states. The deep branch is formed with a sequence of one fully connected layer (DL2), two one-dimensional convolutional layers (CL1 and CL2) each one followed by a max-pooling layer (MP1 and MP2), and finally an additional fully connected layer (DL3). Layer DL2 transforms the input size by increasing the size by a factor of three, acting as a feature enrichment technique. As shown in other works (e.g., \cite{Kravchik2018}), one-dimensional \acp{cnn} are particularly suited for modelling time series data. The purpose of layers CL1 and CL2 is to model the data collected from sensors and actuators in a specific time window. For max pooling, we down-sample the output of each convolutional layer by a factor of 2. The final fully connected layer DL3 re-shapes the output of the deep branch to allow its concatenation with the output of the wide branch.

Both branches are aggregated in another fully connected layer (DL4) followed by multiple dense output sections, each one consisting of the three fully connected layers DL5, DL6 and DL7. Each output section learns the most relevant information from the aggregation layer in order to predict the normal behaviour of a group of sensors controlled by a specific \ac{plc}. 

Each fully connected layer can be described as $Y = LeakyReLU(\mathbf{W}^TX + \mathbf{b})$, where $Y$ is the output, $X$ input, $\mathbf{W}$ is an array of weights the model learns during the training, and $\mathbf{b}$ is the bias. We introduce non-linearity in the model by using the leaky rectified linear activation function defined as follows: $LeakyReLU (x) = max(0, x) + 0.01 * min(0, x)$. Similarly, the convolutional layers can be described as $Y_i = LeakyReLU (Conv(X,\mathbf{W}_i,\mathbf{b}_i))$, where $Y_i$ is the output of the convolution on the input $X$ using the $i$-th filter with weights $\mathbf{W}_i$ and bias $\mathbf{b}_i$.

\textbf{Cost function.} The neural network presented above has been trained to minimise the \ac{mse} cost function by iteratively updating all the weights and biases contained within the feature extractor and all output sections. The cost function computes the error between the prediction of the model and the corresponding observed sensor values. Hence, minimising the cost, we reduce the prediction error. At the training stage, the cost function for a batch size of $s$ samples (i.e., $s$ different time windows) in a specific output section $g$ can be formally written as:

\begin{equation}
\label{eq:mse}
c_{g} = \frac{1}{s}\sum_{t=1}^{s}\Big(\frac{1}{m_{se}^{g}}\sum_{i=1}^{m_{se}^{g}} (Y_t[i] - \tilde{Y_t}[i])^2\Big)
\end{equation}

where $Y_t[i]$ is the predicted value for sensor $i$ at sample $t$ (i.e. time-step $t$), while $\tilde{Y_t}[i]$ is the corresponding observed sensor value (the value present in the training set). $m_{se}^{g}$ is the number of sensors in an output section $g$. The final cost $c$ is the sum of costs of all $G$ output sections $c = \sum_{g=1}^{G} c_{g}$. This cost is minimised by using the optimisation algorithm \ac{sgd} \cite{sutskever2013importance}.  


\subsection{Anomaly detection in actuators}
\label{subsec:anom_det_act}

Actuators in the \ac{swat} and \ac{wadi} testbeds include pumps and motorised valves. The pumps are arranged in pairs of primary and redundant hot-standby pumps. A redundant pump is turned on only in the case the respective primary pump stops working. This operational mode complicates building a forecasting model that predicts the actuator states, as some actuators (such as the redundant pumps), are rarely used during the normal operations. As a consequence, after measuring a high prediction error with \ac{dl}-based methods due to lack of normal operation records, we designed a light and straightforward approach based on querying a database containing all the normal actuator states. The database entries are $n$-tuples, where $n$ is the number of actuators in the testbed ($n=26$ in the \ac{swat}, $n=54$ in the \ac{wadi}). Each entry is a combination of actuators states labelled as normal, for a total of 146 entries available in the \ac{swat} dataset and 2001 entries in the \ac{wadi} dataset. An example of tuple from the \ac{swat} testbed is provided in Table \ref{tab:act_combination}.

\begin{table}[hbp]
	\centering
	\caption{Tuple in the database of actuators normal states. }
	\label{tab:act_combination}
	\begin{tabular}{l|ccccccc}
		\hline
		\textbf{Actuator}&MV101 & P101 & P102 & ... & P601 & P602 & P603  \\ \hline
		\textbf{Tuple}&2     & 2    & 1        & ...    & 1    & 1    & 1     \\ \hline
	\end{tabular}
\end{table}

At testing time, the combinations in the test set with no occurrences in the training set are marked as anomalies, as explained in Section \ref{sec:detection_logic}. 

\subsection{Detection logic}\label{sec:detection_logic} 
\ourtoolext\ operates on batches of actuators and sensors values retrieved from the \acp{plc} at regular time intervals of duration $s$ seconds. For the sake of simplicity, we assume that $s$ also corresponds to the number of readings in one batch for each device (like in the two datasets used in the experiments). We therefore define $I=[\bar{t},\bar{t}+s)$ the time interval under analysis (where $[a,b)$ indicates the interval between $a$ and $b-1$).  Given the observation time $t\in I$, \ourtoolext\ verifies whether the current combination of actuator values is present in the database $A$ built at the training stage. An anomaly is reported otherwise. 

In the case of the sensors, the values observed at time $t$ are compared against those predicted by \ac{wdnn} using past values of sensors and actuators, as previously explained in Section \ref{subsec:anom_sensors}. More precisely, the values observed at time $t\in I$ are compared with the first element of the output of \ac{wdnn} obtained from values of sensors and actuators collected in the time window $[t',t'+W_{in})$ (one input sample for \ac{wdnn}), where $t'=t-H-W_{in}$. The \ac{mse} value of each \ac{wdnn} output section is evaluated against an adaptive threshold to determine the presence of any anomalies. 

\begin{figure}[t!]
	\centering
	\includegraphics[width=\linewidth]{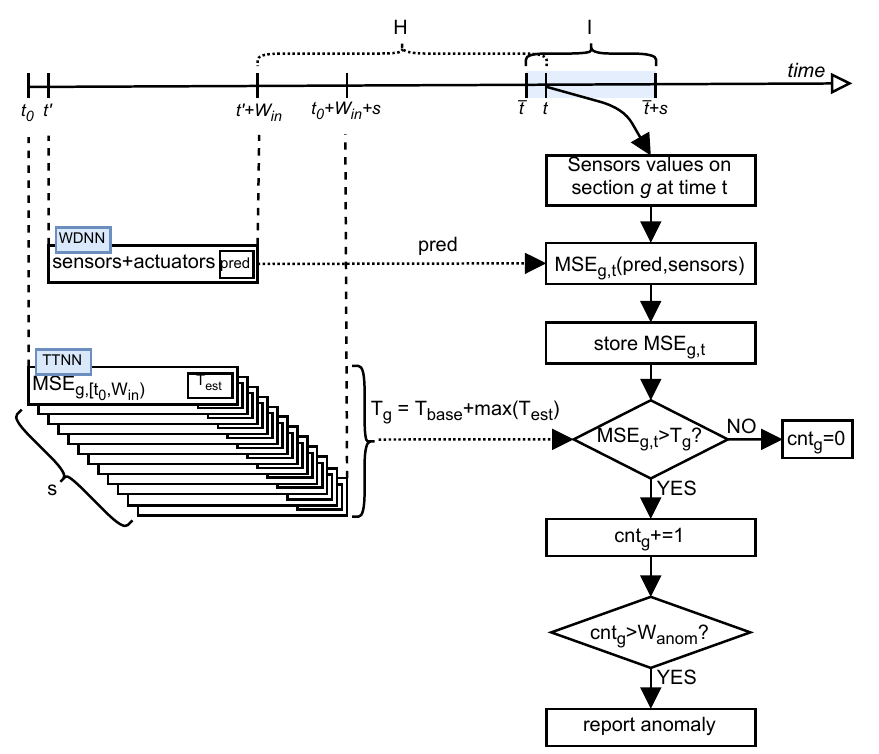}
	\caption{Detection process on section $g$ of sensors through \ac{wdnn} and adaptive thresholds.}
	\label{fig:detection_logic}
\end{figure}

The detection process on one single output section \textit{g} of the neural network is summarised in Figure \ref{fig:detection_logic}, which also depicts an adaptive threshold mechanism for automatically tuning the anomaly threshold based on observed noise conditions on the sensors readings. Such a mechanism is described in Section \ref{subsec:thresh_sel}. The region of the timeline highlighted in light blue represents the time interval $I$ described above.  

More precisely, the samples observed on the group $g$ of sensors at time $t$ are identified as anomalous when the anomaly condition $\ac{mse}_{g,t}>T_g$ is met, where $T_g$ is a threshold on the prediction error for the group of sensors $g$.  Using the same notations as for the cost function in Equation \ref{eq:mse}, we define $\ac{mse}_{g,t}$ as follows:
\begin{equation}
\label{eq:msegt}
\ac{mse}_{g,t} = \frac{1}{m_{se}^{g}}\sum_{i=1}^{m_{se}^{g}} (Y_t[i] - \tilde{Y_t}[i])^2
\end{equation}
To reduce the false positives caused by sudden changes in the underlying physical process (as also reported in \cite{Kravchik2018}), \ourtoolext\ only reports an anomaly at time $t$ on the sensors if the anomaly condition has been also previously observed for $W_{anom}$ consecutive sampling intervals. In Figure \ref{fig:detection_logic}, this process is represented through increasing the counter $cnt_g$.

In summary, the anomaly condition can be expressed as in the following Equation \ref{eq:anomaly_label}.
\begin{equation}
\label{eq:anomaly_label}
L_t=\begin{cases}
1, & \text{if $\exists g\in [1,G]$ s.t.} \\
   & \text{$\ac{mse}_{g,i} > T_g$ $\forall$ $i \in [t - W_{anom}, t]$}\\
1, & \text{if $\nu_{t} \not\in A$}\\
0, & \text{otherwise}
\end{cases}
\end{equation}
where $\nu_{t}$ is the combination of the actuators values observed at time $t$, while $L_t=1$ defines the anomaly condition at time $t$. $W_{anom}$ is one of the hyper-parameters used to tune \ourtoolext, as reported in Section \ref{sec:Results}.

\subsection{Threshold Tuning}
\label{subsec:thresh_sel}

The industrial environments are usually subject to various sources of noise. In particular, the electromagnetic noise can interfere with the communication within the \ac{ics}, hence compromising the operations of anomaly detection systems~\cite{goodwin2008architectures, zhan2012performance}. The main challenge is finding the correct threshold needed to classify an event either as anomaly or as normal activity. Previous works tackled this problem empirically, by tuning the threshold at test time as a hyper-parameter \cite{Shalyga2018, Kravchik2018, goh2017anomaly}. While this technique produces good results in the laboratory when using static datasets such as \ac{swat} and \ac{wadi}, online systems can hardly afford long threshold tuning sessions for updating the thresholds upon new noise levels. Instead, here we propose an adaptive technique that dynamically tunes the threshold based on the prediction error observed previously.    

\begin{figure}[h!]
	\centering
	\includegraphics[width=0.8\linewidth]{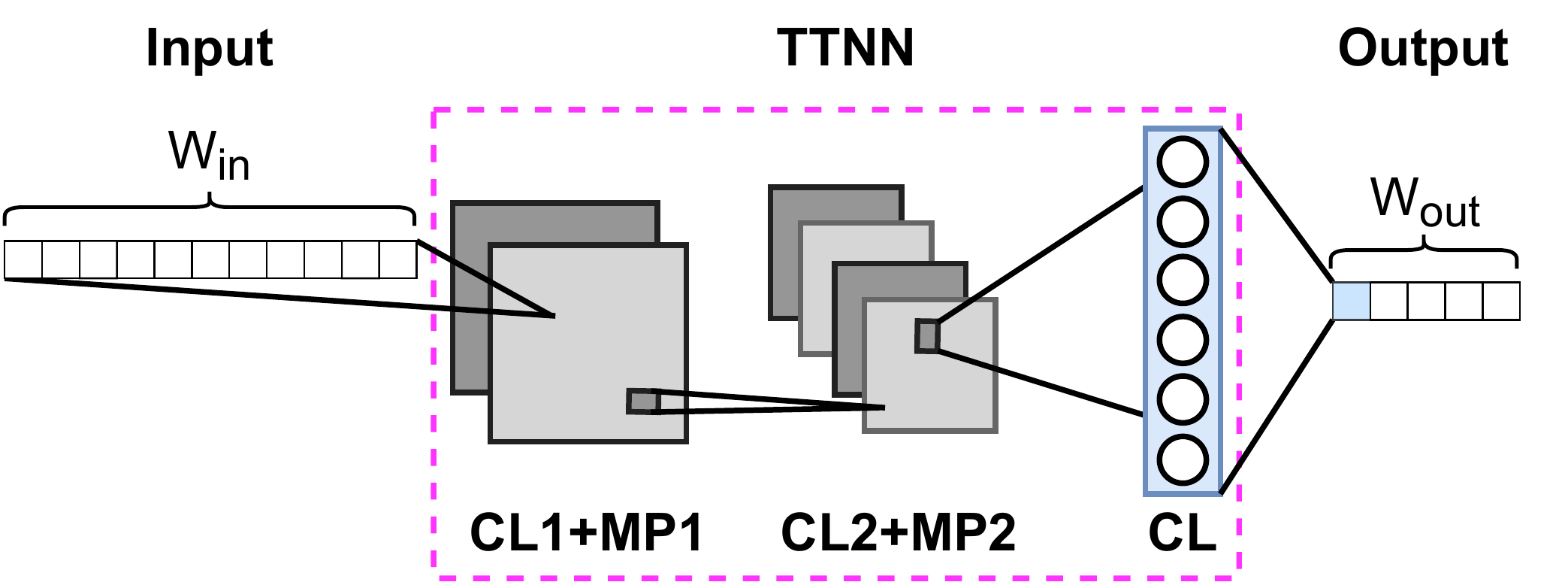}
	\caption{Architecture of \ac{ttnn}. Only the first element of the output is kept. The size of the 1-dim kernel of CL1 and CL2 is 2 with stride 1.}
	\label{fig:arch_aut_threshold}
\end{figure}

Ali et al. \cite{ali2013automated} used a machine learning technique to estimate the threshold based on the historical prediction error of their \ac{ads}. In this paper, we treat the prediction error \ac{mse} as an univariate time series that is modelled using a neural network that we called \ac{ttnn}, whose architecture is depicted in Figure \ref{fig:arch_aut_threshold}. We use $G$ instances of \ac{ttnn}, one for each output section of the \ac{wdnn} model, to tune the thresholds $T_g,\ g\in [1,G]$. Each instance is trained with the prediction error $\ac{mse}_g$ measured on a single output section $g$ on the validation set, which only contains benign records, as the training set. The trained model is used in the online system to compute the optimal anomaly threshold $T_g$ using past prediction errors. As represented in Figure \ref{fig:detection_logic}, the threshold $T_g$ used for anomaly detection on sensors values collected in time interval $[\bar{t},\bar{t}+s)$ is obtained using past prediction errors computed in time interval $[t_0,t_0+s+W_{in})$, where $t_0=\bar{t}-H-W_{in}$.

As shown in Figure \ref{fig:arch_aut_threshold}, \ac{ttnn} consists of two 1-dimensional convolutional layers, each followed by a max pooling layer, and of one final fully connected classification layer, represented in the figure as CL1, MP1, CL2, MP2 and DL respectively. 
The input is a time series of prediction errors measured for the samples collected in past time window of length $W_{in}$ seconds, the output contributes to the estimation of the optimal anomaly threshold for the current sensors states. A detailed description of the threshold tuning process is described in Algorithm \ref{alg:tune_thresh}.      

\begin{algorithm}
	\caption{Threshold tuning algorithm}
	\label{alg:tune_thresh}
	\begin{algorithmic}[1]
		\renewcommand{\algorithmicrequire}{\textbf{Input:}}
		\renewcommand{\algorithmicensure}{\textbf{Output:}}
		\Require Batch of past prediction errors on section $g$ ($E_g = \{\ac{mse}_{g,[t_0,t_0+W_{in})}$,...,$\ac{mse}_{g,[t_0+s,t_0+W_{in}+s)}\}$), where $t_0=\bar{t}-H-W_{in}$
		\Ensure Anomaly threshold for \ac{wdnn} output section $g$ ($T_g$)
		\State $T_{est} =$ []; \Comment{List of estimated thresholds}
		\State $\ac{ttnn}[g]\gets $ load the weights and biases for section $g$;
		\State $\bar{E}_g\gets$ median($E_g$) \Comment{Median filter on the input}
		\For{$X = \bar{\ac{mse}}_{g,[t_0+i,t_0+W_{in}+i)}\in\bar{E}_g$ s.t. $i\in[0,s)$} 
		\State $\hat{Y} = \ac{ttnn}[g](X)$; \Comment{Estimated prediction error}
		\State $T_{est}.append(\hat{Y}[0])$;
		\EndFor
		\State $T_g = T_{base} + max(T_{est})$;
	\end{algorithmic}
\end{algorithm}

Referring to the timeline depicted in Figure \ref{fig:detection_logic}, the anomaly threshold used at time $t\in[\bar{t},\bar{t}+s)$ for output section $g$ is computed by using a batch $E_g$ of $s$ samples defined as time series of past prediction errors of length $W_{in}$, starting from $\ac{mse}_{g,[t_0,t_0+W_{in})}$,  $\ac{mse}_{g,[t_0+1,t_0+W_{in}+1)}$, to $\ac{mse}_{g,[t_0+s,t_0+W_{in}+s)}$. 

At line 3 in of the algorithm, a median filter is applied to the input batch of past prediction errors $E_g$ to reduce the impact of short-term changes of the prediction error on the threshold tuning process. Median filtering is accomplished by sliding a window over $E_g$ while computing the median value of the elements of $E_g$ under the window. The resulting median values are stored in array $\bar{E}_g$. One variable to consider in this process is the size of the median filter. In this regard, we experimented with values between 50 and 120. As reported in Table \ref{tab:hyperparameters}, we obtained the best results with median filter size of $59$. 
In the loop at lines 4-7, the algorithm computes the estimated prediction error using the median values in $\bar{E}_g$. As shown at line 6, only the first element of the output layer is memorised for further processing. 

The threshold $T_g$ is computed at line 8 as the maximum of the predicted thresholds plus the offset $T_{base}$. $T_{base}$ is set as the sum of the mean plus the standard deviation of the prediction error on the validation set, and is kept constant after the deployment. This offset is added to reduce the sensitivity of the \ac{wdnn} to small variations on sensors values, hence reducing the false positives.

As a final step, the algorithm executes one \ac{sgd} epoch to update weights and biases of each of the $G$ instances of \ac{ttnn} using the respective input batch $E_g$.

\subsection{The few-time-steps algorithm}\label{sec:few-time-steps}
The few-time-steps algorithm has been designed to efficiently reconfigure \ourtoolext\ in a production environment in the case an anomaly is identified by the system and then recognised as false alarm by the technician.

In such a situation, \ourtoolext\ is triggered by the technician to update the database $A$ of actuators states and to tune the output section of the pre-trained \ac{wdnn}. The only assumption is that the technician can recognise false alarms caused by changes in the normal operating condition of the \ac{ics} (e.g., changes in the hardware/software configurations). Unlike state-of-the-art solutions, the proposed approach only assumes a domain-specific understanding of the \ac{ics} operations, without requiring a deep knowledge of the algorithms and their thresholds and hyper-parameters.

Of course, the number of human interventions needed to handle the false alarms is an important parameter to determine the usability of the anomaly detection system. If the system is too sensitive, the technician would be overwhelmed by a large number of alarms to be verified, making it hard to spot those that are due to normal changes in the \ac{ics} behaviour, hence false. On the contrary, a conservative approach might not reveal true malicious activities or anomalies. Both cases can render the system ineffective, if not unusable. The sensitivity of \ourtoolext\ to this parameter will be analysed in Section \ref{sec:Results}.
\begin{algorithm}
	\caption{Few-time-steps learning algorithm}
	\label{alg:few-time-steps}
	\begin{algorithmic}[1]
		\renewcommand{\algorithmicrequire}{\textbf{Input:}}
		\renewcommand{\algorithmicensure}{\textbf{Output:}}
		\Require Batch of samples ($S$), Technician's input: False alarm at time $t$ ($FA_t[i]$) $i\in[0,G]$
		\Ensure Retrained output section  ($\mu_{out}[g]$)
		\If{$\nu_t\not\in A$ \textbf{and} $FA_t[0] ==$ \textbf{True}} 
		\State $A = A \cup \nu_t$; \Comment{Update the database of actuators}
		\EndIf
		\State $EP_{tuning} \gets$ number of fine-tuning epochs;
		\For {$g\in[1,G]$}
		\If {$FA_t[g] ==$ \textbf{True}} 
		\State $\mu_{out}[g] \gets $ weights and biases
		\For {epoch \textbf{in} $EP_{tuning}$}
		\State $c_g \gets$ cost of section $g$ on $S$;
		\State $SGD(\mu_{out}[g])$ step to minimise $c_g$;
		\State $\mu_{out}[g] \gets $ update weights and biases;
		\EndFor
		\EndIf
		\EndFor
	\end{algorithmic}
\end{algorithm}

Algorithm \ref{alg:few-time-steps} is called upon an technician's decision that an alarm detected by the system at time $t$ is false. This means that Algorithm \ref{alg:few-time-steps} is called when both of the following conditions are satisfied: the anomaly condition $L_t=1$ and at least one of the elements of the boolean vector $FA_t[i]$ ($i\in[0,G]$) is \textit{True}. $FA_t$ indicates where the false alarm has been detected, either among the actuators ($FA_t[0] = True$), or in one or more groups of sensors ($FA_t[i]=True,\ i\in[1,G]$), or both.

In the first case, with $FA_t[0] = True$, the algorithm adds the combination of actuator states $\nu_t$ to database $A$ (lines 1-3), as the technician has determined that $\nu_t$ is normal and must be treated as such by the system. In lines 5-14, the output sections of the neural network that have produced the false alarm are updated. Specifically, in lines 8-12, \ac{sgd} is employed to fine-tune the output section through multiple gradient steps. We calculate the prediction loss for the data samples aggregated in a \textit{batch} of $S$ samples containing the false alarm ($S$ is passed as an input to algorithm). The optimiser minimises this loss by tuning the parameters of the output layers DL, DL6, and DL7. After $EP_{tuning}$ optimisation steps (around 400 \textit{ms} on average for $100$ epochs on our testing environment described in Section \ref{sec:Experimental_setup}), the updated output section $\mu_{out}[g]$ replaces the previous one in the anomaly detection process (line 11). Note that, grouping the sensors into different sections speeds up the execution of the few time steps algorithm, since only a portion of the output section is updated in the case of false alarms.

\section{Experimental evaluation}\label{sec:Results}
\input{results_data}
\input{results_data}
\newcommand{\chartWADINoiseTestMedfilterFSText}{
	\begin{tikzpicture}
	\begin{axis}[  
		height=4 cm,
		width=7 cm,
		font=\footnotesize,
		grid = both,
		xlabel={Noise $\sigma$},
		ylabel={F1-score},
		legend style={nodes={scale=0.6}, fill opacity=0.6, text opacity=1, at={(0.5,0.45))},anchor=north},
		legend columns=2,
		scaled y ticks=false,
		scaled x ticks=false,
		xmin=0,
		xmax=15,
		xtick={0,1,2,3,5,10,15},
		xticklabels={0,1,2,3,5,10,15},
		xtick pos=left,
		ymin=0.3, ymax=0.9,
		ytick={0.3,0.4,0.5,0.6,0.7,0.8,0.9},
		yticklabels={0.3,0.4,0.5,0.6,0.7,0.8,0.9},
		y label style={at={(axis description cs:-0.15,.5)}},
		ytick pos=left,
		enlargelimits=0.02, 
		]
		\addplot [color=blue,style= thick,mark=*,mark size=1.5] table[x index=0,y index=1] {\WADINoiseTestMedfilterFS};\addlegendentry{$med\_kernel=51$}
		\addplot [color=orange,style= thick,mark=triangle*,mark size=1.5] table[x index=0,y index=2] {\WADINoiseTestMedfilterFS}; \addlegendentry{$med\_kernel=53$}
		\addplot [color=green!50!black,style= thick,mark=square*,mark size=1.5] table[x index=0,y index=3] {\WADINoiseTestMedfilterFS}; \addlegendentry{$med\_kernel=55$}
		\addplot [color=purple!80,style= thick,mark=diamond*,mark size=2] table[x index=0,y index=4] {\WADINoiseTestMedfilterFS}; \addlegendentry{$med\_kernel=57$}
		\addplot [color=brown!85!black,style= thick,mark=star,mark size=2] table[x index=0,y index=5] {\WADINoiseTestMedfilterFS}; \addlegendentry{$med\_kernel=59$}
	\end{axis}
	\end{tikzpicture}
}

\newcommand{\chartWADINoiseTestMedfilterAttText}{
	\begin{tikzpicture}
	\begin{axis}[  
		height=4 cm,
		width=7 cm,
		font=\footnotesize,
		grid = both,
		xlabel={Noise $\sigma$},
		legend style={nodes={scale=0.6}, fill opacity=0.6, text opacity=1, at={(0.5,0.45))},anchor=north},
		legend columns=2,
		scaled y ticks=false,
		scaled x ticks=false,
		xmin=0,
		xmax=15,
		xtick={0,1,2,3,5,10,15},
		xticklabels={0,1,2,3,5,10,15},
		xtick pos=left,
		ylabel={\# of detected attacks},
		ymin=3, ymax=15,
		ytick={3,4,5,6,7,8,9,10,11,12,13,14,15},
		yticklabels={3,4,5,6,7,8,9,10,11,12,13,14,15},
		y label style={at={(axis description cs:-0.15,.5)}},
		ytick pos=left,
		enlargelimits=0.02, 
		]
		\addplot [color=blue,style= thick,mark=*,mark size=1.5] table[x index=0,y index=1] {\WADINoiseTestMedfilterAtt};\addlegendentry{$med\_kernel=51$}
		\addplot [color=orange,style= thick,mark=triangle*,mark size=1.5] table[x index=0,y index=2] {\WADINoiseTestMedfilterAtt}; \addlegendentry{$med\_kernel=53$}
		\addplot [color=green!50!black,style= thick,mark=square*,mark size=1.5] table[x index=0,y index=3] {\WADINoiseTestMedfilterAtt}; \addlegendentry{$med\_kernel=55$}
		\addplot [color=purple!80,style= thick,mark=diamond*,mark size=2] table[x index=0,y index=4] {\WADINoiseTestMedfilterAtt}; \addlegendentry{$med\_kernel=57$}
		\addplot [color=brown!85!black,style= thick,mark=star,mark size=2] table[x index=0,y index=5] {\WADINoiseTestMedfilterAtt}; \addlegendentry{$med\_kernel=59$}
	\end{axis}
	\end{tikzpicture}
}

\newcommand{\chartWADIWAnomalyText}{
	\begin{tikzpicture}
	\begin{axis}[  
		axis y line*=left,
		height=4.5 cm,
		width=0.9\linewidth,
		font=\tiny,
		grid = both,
		xlabel={$W_{anom}$ (sec)},
		ylabel={Performance metric},
		scaled y ticks=false,
		scaled x ticks=false,
		xmin=27,
		xmax=39,
		xtick={27, 30, 33, 36, 39},
		xticklabels={27, 30, 33, 36, 39},
		xtick pos=left,
		ymin=0.5, ymax=1.0,
		ytick={0.5,0.6,0.7,0.8,0.9, 1.0},
		yticklabels={0.5,0.6,0.7,0.8,0.9, 1.0},
		ytick pos=left,
		enlargelimits=0.02, 
		]
		\addplot [color=red,style= thick,mark=*,mark size=2,mark options={solid}] table[x index=0,y index=1] {\WADIWAnomaly}; \label{WADIWAnomaly_prec}
		\addplot [color=green!70!black,style= thick,mark=square*,mark size=2,mark options={solid}] table[x index=0,y index=2] {\WADIWAnomaly}; \label{WADIWAnomaly_recall}
		\addplot [color=blue,style= thick,mark=triangle*,mark size=2.5,mark options={solid}] table[x index=0,y index=3] {\WADIWAnomaly}; \label{WADIWAnomaly_f1}
	\end{axis}
	
	\begin{axis}[  
		axis y line*=right,
		axis x line=none,
		legend style={nodes={scale=0.6}, fill opacity=0.6, text opacity=1},
		legend columns=2,
		legend pos = south east,
		height=4.5 cm,
		width=0.9\linewidth,
		font=\tiny,
		grid = major,
		ylabel={\# of detected attacks},
		ymin=7, ymax=17,
		ytick={7,8,9,10,11,12,13,14,15, 16, 17},
		yticklabels={7,8,9,10,11,12,13,14,15, 16, 17},
		ytick pos=right,
		enlargelimits=0.02,
	]
	\addlegendimage{/pgfplots/refstyle=WADIWAnomaly_prec}\addlegendentry{Precision}
	\addlegendimage{/pgfplots/refstyle=WADIWAnomaly_recall}\addlegendentry{Recall}
	\addlegendimage{/pgfplots/refstyle=WADIWAnomaly_f1}\addlegendentry{F1-score}
	\addplot [color=orange,dashed,style= thick,mark=star,mark size=3,mark options={solid}] table[x index=0,y index=4] {\WADIWAnomaly};
	\addlegendentry{\# attacks}
	\end{axis}
	\end{tikzpicture}
}

\newcommand{\chartWADIMedFilterText}{
	\begin{tikzpicture}
	\begin{axis}[  
		axis y line*=left,
		height=4 cm,
		width=4.6 cm,
		font=\footnotesize,
		grid = both,
		xlabel={$med\_kernel$},
		ylabel={Performance metric},
		scaled y ticks=false,
		scaled x ticks=false,
		xmin=53,
		xmax=61,
		xtick={53, 55, 57, 59, 61},
		xticklabels={53, 55, 57, 59, 61},
		xtick pos=left,
		ymin=0.4, ymax=1.0,
		ytick={0.4,0.5,0.6,0.7,0.8,0.9, 1.0},
		yticklabels={0.4,0.5,0.6,0.7,0.8,0.9, 1.0},
		y label style={at={(axis description cs:-0.15,.5)}},
		ytick pos=left,
		enlargelimits=0.02, 
		]
		\addplot [color=blue,style= thick,mark=*,mark size=1.5] table[x index=0,y index=1] {\WADIMedFilter}; \label{WADIMedFilter_prec}
		\addplot [color=orange,style= thick,mark=triangle*,mark size=1.5] table[x index=0,y index=2] {\WADIMedFilter}; \label{WADIMedFilter_recall}
		\addplot [color=green!50!black,style= thick,mark=square*,mark size=1.5] table[x index=0,y index=3] {\WADIMedFilter}; \label{WADIMedFilter_f1}
	\end{axis}
	
	\begin{axis}[  
		axis y line*=right,
		axis x line=none,
		legend style={nodes={scale=0.6}, fill opacity=0.6, text opacity=1, at={(0.5,0.35))},anchor=north},
		legend columns=2,
		height=4 cm,
		width=4.6 cm,
		font=\footnotesize,
		grid = major,
		ylabel={\# of detected attacks},
		ymin=3, ymax=15,
		ytick={3,4,5,6,7,8,9,10,11,12,13,14,15},
		yticklabels={3,4,5,6,7,8,9,10,11,12,13,14,15},
		y label style={at={(axis description cs:1.1,.5)}},
		ytick pos=right,
		enlargelimits=0.02,
	]
	\addlegendimage{/pgfplots/refstyle=WADIMedFilter_prec}\addlegendentry{Precision}
	\addlegendimage{/pgfplots/refstyle=WADIMedFilter_recall}\addlegendentry{Recall}
	\addlegendimage{/pgfplots/refstyle=WADIMedFilter_f1}\addlegendentry{F1-score}
	\addplot [color=magenta!80,dashed,style= thick,mark=*,mark size=1.5] table[x index=0,y index=4] {\WADIMedFilter};
	\addlegendentry{\# attacks}
	\end{axis}
	\end{tikzpicture}
}

\newcommand{\chartWADIHIntMedFilterText}{
	\begin{tikzpicture}
	\begin{axis}[  
		axis y line*=left,
		height=4 cm,
		width=4.6 cm,
		font=\footnotesize,
		grid = both,
		xlabel={$med\_kernel$},
		ylabel={Performance metric},
		scaled y ticks=false,
		scaled x ticks=false,
		xmin=53,
		xmax=61,
		xtick={53, 55, 57, 59, 61},
		xticklabels={53, 55, 57, 59, 61},
		xtick pos=left,
		ymin=0.4, ymax=1.0,
		ytick={0.4,0.5,0.6,0.7,0.8,0.9, 1.0},
		yticklabels={0.4,0.5,0.6,0.7,0.8,0.9, 1.0},
		y label style={at={(axis description cs:-0.15,.5)}},
		ytick pos=left,
		enlargelimits=0.02, 
		]
		\addplot [color=blue,style= thick,mark=*,mark size=1.5] table[x index=0,y index=1] {\WADIHIntMedFilter}; \label{WADIHIntMedFilter_prec}
		\addplot [color=orange,style= thick,mark=triangle*,mark size=1.5] table[x index=0,y index=2] {\WADIHIntMedFilter}; \label{WADIHIntMedFilter_recall}
		\addplot [color=green!50!black,style= thick,mark=square*,mark size=1.5] table[x index=0,y index=3] {\WADIHIntMedFilter}; \label{WADIHIntMedFilter_f1}
	\end{axis}
	
	\begin{axis}[  
		axis y line*=right,
		axis x line=none,
		legend style={nodes={scale=0.47}, fill opacity=0.6, text opacity=1, at={(0.5,0.3))},anchor=north},
		legend columns=2,
		height=4 cm,
		width=4.6 cm,
		font=\footnotesize,
		grid = major,
		ylabel={\# of human interventions},
		ymin=180, ymax=300,
		ytick={180, 190, 200, 210, 220, 230, 240, 250, 260, 270, 280, 290, 300},
		yticklabels={180, 190, 200, 210, 220, 230, 240, 250, 260, 270, 280, 290, 300},
		y label style={at={(axis description cs:1.15,.5)}},
		ytick pos=right,
		enlargelimits=0.02,
	]
	\addlegendimage{/pgfplots/refstyle=WADIHIntMedFilter_prec}\addlegendentry{Precision}
	\addlegendimage{/pgfplots/refstyle=WADIHIntMedFilter_recall}\addlegendentry{Recall}
	\addlegendimage{/pgfplots/refstyle=WADIHIntMedFilter_f1}\addlegendentry{F1-score}
	\addplot [color=purple!80,dashed,style= thick,mark=*,mark size=1.5] table[x index=0,y index=4] {\WADIHIntMedFilter};
	\addlegendentry{\# Interventions}
	\end{axis}
	\end{tikzpicture}
}

\newcommand{\chartWADIHIntWSilentText}{
	\begin{tikzpicture}
	\begin{axis}[  
		axis y line*=left,
		height=4.5 cm,
		width=0.9\linewidth,
		grid = both,
		xlabel={$W_{grace}$ (sec)},
		ylabel={Performance metric},
		scaled y ticks=false,
		scaled x ticks=false,
		xmin=0,
		xmax=20,
		xtick={0, 5, 10, 15, 20},
		xticklabels={0, 5, 10, 15, 20},
		xtick pos=left,
		ymin=0.5, ymax=1.0,
		ytick={0.5,0.6,0.7,0.8,0.9, 1.0},
		yticklabels={0.5,0.6,0.7,0.8,0.9, 1.0},
		ytick pos=left,
		enlargelimits=0.02, 
		]
		\addplot [color=red,style= thick,mark=*,mark size=2,mark options={solid}] table[x index=0,y index=1] {\WADIHIntWSilent}; \label{WADIHIntWSilent_prec}
		\addplot [color=green!70!black,style= thick,mark=square*,mark size=2,mark options={solid}] table[x index=0,y index=2] {\WADIHIntWSilent}; \label{WADIHIntWSilent_recall}
		\addplot [color=blue,style= thick,mark=triangle*,mark size=2.5,mark options={solid}] table[x index=0,y index=3] {\WADIHIntWSilent}; \label{WADIHIntWSilent_f1}
	\end{axis}
	
	\begin{axis}[  
		axis y line*=right,
		axis x line=none,
		legend style={nodes={scale=0.6}, fill opacity=0.6, text opacity=1},
		legend columns=2,
		legend pos = north east,
		height=4.5 cm,
		width=0.9\linewidth,
		grid = major,
		ylabel style={align=center}, ylabel={Average \# of\\ interventions per hour},
		ymin=0, ymax=10,
		ytick={0,1,2,3,4,5,6,7,8,9,10},
		yticklabels={0,1,2,3,4,5,6,7,8,9,10},
		ytick pos=right,
		enlargelimits=0.02,
	]
	\addlegendimage{/pgfplots/refstyle=WADIHIntWSilent_prec}\addlegendentry{Precision}
	\addlegendimage{/pgfplots/refstyle=WADIHIntWSilent_recall}\addlegendentry{Recall}
	\addlegendimage{/pgfplots/refstyle=WADIHIntWSilent_f1}\addlegendentry{F1-score}
	\addplot [color=orange,dashed,style= thick,mark=star,mark size=3,mark options={solid}] table[x index=0,y index=4] {\WADIHIntWSilent};
	\addlegendentry{\# Interventions}
	\end{axis}
	\end{tikzpicture}
}

\newcommand{\chartSWATWAnomalyText}{
	\begin{tikzpicture}
	\begin{axis}[  
		axis y line*=left,
		height=4.5 cm,
		width=0.9\linewidth,
		font=\tiny,
		grid = both,
		xlabel={$W_{anom}$ (sec)},
		ylabel={Performance metric},
		scaled y ticks=false,
		scaled x ticks=false,
		xmin=27,
		xmax=39,
		xtick={27, 30, 33, 36, 39},
		xticklabels={27, 30, 33, 36, 39},
		xtick pos=left,
		ymin=0.5, ymax=1,
		ytick={0.5,0.6,0.7,0.8,0.9,1.0},
		yticklabels={0.5,0.6,0.7,0.8,0.9,1.0},
		ytick pos=left,
		enlargelimits=0.02, 
		]
		
		\addplot [color=red,style= thick,mark=*,mark size=2,mark options={solid}] table[x index=0,y index=1] {\SWATWAnomaly}; \label{SWATWAnomaly_prec}
		\addplot [color=green!70!black,style= thick,mark=square*,mark size=2,mark options={solid}] table[x index=0,y index=2] {\SWATWAnomaly}; \label{SWATWAnomaly_recall}
		\addplot [color=blue,style= thick,mark=triangle*,mark size=2.5,mark options={solid}] table[x index=0,y index=3] {\SWATWAnomaly}; \label{SWATWAnomaly_f1}
	\end{axis}
	
	\begin{axis}[  
		axis y line*=right,
		axis x line=none,
		legend style={nodes={scale=0.6}, fill opacity=0.6, text opacity=1},
		legend columns=2,
		legend pos = south east,
		height=4.5 cm,
		width=0.9\linewidth,
		font=\tiny,
		grid = major,
		ylabel={\# of detected attacks},
		ymin=28, ymax=38,
		ytick={28,29,30,31,32,33,34,35,36,37,38},
		yticklabels={28,29,30,31,32,33,34,35,36,37,38},
		ytick pos=right,
		enlargelimits=0.02,
	]
	\addlegendimage{/pgfplots/refstyle=SWATWAnomaly_prec}\addlegendentry{Precision}
	\addlegendimage{/pgfplots/refstyle=SWATWAnomaly_recall}\addlegendentry{Recall}
	\addlegendimage{/pgfplots/refstyle=SWATWAnomaly_f1}\addlegendentry{F1-score}
	\addplot [color=orange,dashed,style= thick,mark=star,mark size=3,mark options={solid}] table[x index=0,y index=4] {\SWATWAnomaly};
	\addlegendentry{\# attacks}
	\end{axis}
	\end{tikzpicture}
}

\newcommand{\chartSWATMedFilterText}{
	\begin{tikzpicture}
	\begin{axis}[  
		axis y line*=left,
		height=4 cm,
		width=4.6 cm,
		font=\footnotesize,
		grid = both,
		xlabel={$med\_kernel$},
		ylabel={Performance metric},
		scaled y ticks=false,
		scaled x ticks=false,
		xmin=53,
		xmax=61,
		xtick={53, 55, 57, 59, 61},
		xticklabels={53, 55, 57, 59, 61},
		xtick pos=left,
		ymin=0.4, ymax=1.0,
		ytick={0.4,0.5,0.6,0.7,0.8,0.9, 1.0},
		yticklabels={0.4,0.5,0.6,0.7,0.8,0.9, 1.0},
		y label style={at={(axis description cs:-0.15,.5)}},
		ytick pos=left,
		enlargelimits=0.02, 
		]
		\addplot [color=blue,style= thick,mark=*,mark size=1.5] table[x index=0,y index=1] {\SWATMedFilter}; \label{SWATMedFilter_prec}
		\addplot [color=orange,style= thick,mark=triangle*,mark size=1.5] table[x index=0,y index=2] {\SWATMedFilter}; \label{SWATMedFilter_recall}
		\addplot [color=green!50!black,style= thick,mark=square*,mark size=1.5] table[x index=0,y index=3] {\SWATMedFilter}; \label{SWATMedFilter_f1}
	\end{axis}
	
	\begin{axis}[  
		axis y line*=right,
		axis x line=none,
		legend style={nodes={scale=0.6}, fill opacity=0.6, text opacity=1, at={(0.5,0.35))},anchor=north},
		legend columns=2,
		height=4 cm,
		width=4.6 cm,
		font=\footnotesize,
		grid = major,
		ylabel={\# of detected attacks},
		ymin=26, ymax=38,
		ytick={26,27,28,29,30,31,32,33,34,35,36,37,38},
		yticklabels={26,27,28,29,30,31,32,33,34,35,36,37,38},
		y label style={at={(axis description cs:1.1,.5)}},
		ytick pos=right,
		enlargelimits=0.02,
	]
	\addlegendimage{/pgfplots/refstyle=SWATMedFilter_prec}\addlegendentry{Precision}
	\addlegendimage{/pgfplots/refstyle=SWATMedFilter_recall}\addlegendentry{Recall}
	\addlegendimage{/pgfplots/refstyle=SWATMedFilter_f1}\addlegendentry{F1-score}
	\addplot [color=magenta!80,dashed,style= thick,mark=*,mark size=1.5] table[x index=0,y index=4] {\SWATMedFilter};
	\addlegendentry{\# attacks}
	\end{axis}
	\end{tikzpicture}
}

\newcommand{\chartSWATHIntMedFilterText}{
	\begin{tikzpicture}
	\begin{axis}[  
		axis y line*=left,
		height=4 cm,
		width=4.6 cm,
		font=\footnotesize,
		grid = both,
		xlabel={$med\_kernel$},
		ylabel={Performance metric},
		scaled y ticks=false,
		scaled x ticks=false,
		xmin=53,
		xmax=61,
		xtick={53, 55, 57, 59, 61},
		xticklabels={53, 55, 57, 59, 61},
		xtick pos=left,
		ymin=0.4, ymax=1.0,
		ytick={0.4,0.5,0.6,0.7,0.8,0.9, 1.0},
		yticklabels={0.4,0.5,0.6,0.7,0.8,0.9, 1.0},
		y label style={at={(axis description cs:-0.15,.5)}},
		ytick pos=left,
		enlargelimits=0.02, 
		]
		\addplot [color=blue,style= thick,mark=*,mark size=1.5] table[x index=0,y index=1] {\SWATHIntMedFilter}; \label{SWATHIntMedFilter_prec}
		\addplot [color=orange,style= thick,mark=triangle*,mark size=1.5] table[x index=0,y index=2] {\SWATHIntMedFilter}; \label{SWATHIntMedFilter_recall}
		\addplot [color=green!50!black,style= thick,mark=square*,mark size=1.5] table[x index=0,y index=3] {\SWATHIntMedFilter}; \label{SWATHIntMedFilter_f1}
	\end{axis}
	
	\begin{axis}[  
		axis y line*=right,
		axis x line=none,
		legend style={nodes={scale=0.47}, fill opacity=0.6, text opacity=1, at={(0.5,0.5))},anchor=north},
		legend columns=2,
		height=4 cm,
		width=4.6 cm,
		font=\footnotesize,
		grid = major,
		ylabel={\# of human interventions},
		ymin=0, ymax=900,
		ytick={0,75, 150, 225, 300, 375, 450, 525, 600, 675, 750, 825,900},
		yticklabels={0,75, 150, 225, 300, 375, 450, 525, 600, 675, 750, 825,900},
		y label style={at={(axis description cs:1.15,.5)}},
		ytick pos=right,
		enlargelimits=0.02,
	]
	\addlegendimage{/pgfplots/refstyle=SWATHIntMedFilter_prec}\addlegendentry{Precision}
	\addlegendimage{/pgfplots/refstyle=SWATHIntMedFilter_recall}\addlegendentry{Recall}
	\addlegendimage{/pgfplots/refstyle=SWATHIntMedFilter_f1}\addlegendentry{F1-score}
	\addplot [color=purple!80,dashed,style= thick,mark=*,mark size=1.5] table[x index=0,y index=4] {\SWATHIntMedFilter};
	\addlegendentry{\# Interventions}
	\end{axis}
	\end{tikzpicture}
}

\newcommand{\chartSWATHIntWSilentText}{
	\begin{tikzpicture}
	\begin{axis}[  
		axis y line*=left,
		height=4.5 cm,
		width=0.9\linewidth,
		grid = both,
		xlabel={$W_{grace}$ (sec)},
		ylabel={Performance metric},
		scaled y ticks=false,
		scaled x ticks=false,
		xmin=0,
		xmax=20,
		xtick={0, 5, 10, 15, 20},
		xticklabels={0, 5, 10, 15, 20},
		xtick pos=left,
		ymin=0.5, ymax=1.0,
		ytick={0.5,0.6,0.7,0.8,0.9, 1.0},
		yticklabels={0.5,0.6,0.7,0.8,0.9, 1.0},
		ytick pos=left,
		enlargelimits=0.02, 
		]
		\addplot [color=red,style= thick,mark=*,mark size=2,mark options={solid}] table[x index=0,y index=1] {\SWATHIntWSilent}; \label{SWATHIntWSilent_prec}
		\addplot [color=green!70!black,style= thick,mark=square*,mark size=2,mark options={solid}] table[x index=0,y index=2] {\SWATHIntWSilent}; \label{SWATHIntWSilent_recall}
		\addplot [color=blue,style= thick,mark=triangle*,mark size=2.5,mark options={solid}] table[x index=0,y index=3] {\SWATHIntWSilent}; \label{SWATHIntWSilent_f1}
	\end{axis}
	
	\begin{axis}[  
		axis y line*=right,
		axis x line=none,
		legend style={nodes={scale=0.6}, fill opacity=0.6, text opacity=1, at={(0.72,0.5))},anchor=north},
		legend columns=2,
		height=4.5 cm,
		width=0.9\linewidth,
		grid = major,
		ylabel style={align=center}, ylabel={Average \# of\\ interventions per hour},
		ymin=0, ymax=10,
		ytick={0,1,2,3,4,5,6,7,8,9,10},
		yticklabels={0,1,2,3,4,5,6,7,8,9,10},
		ytick pos=right,
		enlargelimits=0.02,
	]
	\addlegendimage{/pgfplots/refstyle=SWATHIntWSilent_prec}\addlegendentry{Precision}
	\addlegendimage{/pgfplots/refstyle=SWATHIntWSilent_recall}\addlegendentry{Recall}
	\addlegendimage{/pgfplots/refstyle=SWATHIntWSilent_f1}\addlegendentry{F1-score}
	\addplot [color=orange,dashed,style= thick,mark=star,mark size=3,mark options={solid}] table[x index=0,y index=4] {\SWATHIntWSilent};
	\addlegendentry{\# Interventions}
	\end{axis}
	\end{tikzpicture}
}

\newcommand{\chartSWATNoiseTestMedfilterFSText}{
	\begin{tikzpicture}
	\begin{axis}[  
	axis y line*=left,
	height=4.5 cm,
	width=8 cm,
	grid = both,
	xlabel={Noise $\sigma$},
	ylabel={F1 score},
	scaled y ticks=false,
	scaled x ticks=false,
	xmin=0,
	xmax=15,
	xtick={0,1,2,3,5,10,15},
	xticklabels={0,1,2,3,5,10,15},
	xtick pos=left,
	ymin=0.3, ymax=1.0,
	ytick={0.3,0.4,0.5,0.6,0.7,0.8,0.9, 1.0},
	yticklabels={0.3,0.4,0.5,0.6,0.7,0.8,0.9, 1.0},
	ytick pos=left,
	enlargelimits=0.02, 
	]
	\addplot [color=blue,style= thick,mark=*,mark size=2] table[x index=0,y index=5] {\SWATNoiseTestMedfilterFS}; \label{SWATNoise_FS}
	\end{axis}
	
	\begin{axis}[  
	axis y line*=right,
	axis x line=none,
	legend style={nodes={scale=0.6}, fill opacity=0.6},
	legend columns=2,
	legend pos = south east,
	height=4.5 cm,
	width=8 cm,
	grid = major,
	ylabel={\# of detected attacks},
	ymin=20, ymax=34,
	ytick={20,22,24,26,28,30,32,34},
	yticklabels={20,22,24,26,28,30,32,34},
	ytick pos=right,
	enlargelimits=0.02,
	]
	\addlegendimage{/pgfplots/refstyle=SWATNoise_FS}\addlegendentry{F1 score}
	\addplot [color=green!50!black,style= thick,mark=square*,mark size=2] table[x index=0,y index=5] {\SWATNoiseTestMedfilterAtt}; \addlegendentry{\# of detected attacks}
	\end{axis}
	\end{tikzpicture}
}

\newcommand{\chartSWATNoise}{
	\begin{tikzpicture}
	\begin{axis}[  
	axis y line*=left,
	height=4.5 cm,
	width=0.9\linewidth,
	font=\tiny,
	grid = both,
	xlabel={Noise $\sigma$},
	ylabel={Performance Metric},
	scaled y ticks=false,
	scaled x ticks=false,
	xmin=0,
	xmax=15,
	xtick={0,1,2,3,5,10,15},
	xticklabels={0,1,2,3,5,10,15},
	xtick pos=left,
	ymin=0.5, ymax=1,
	ytick={0.5,0.6,0.7,0.8,0.9,1.0},
	yticklabels={0.5,0.6,0.7,0.8,0.9,1.0},
	ytick pos=left,
	enlargelimits=0.02, 
	]
	\addplot [color=red,style= thick,mark=*,mark size=2,mark options={solid}] table[x index=0,y index=1] {\SWATNoiseTest}; \label{SWATWAnomaly_p}
	\addplot [color=green!70!black,style= thick,mark=square*,mark size=2,mark options={solid}] table[x index=0,y index=2] {\SWATNoiseTest}; \label{SWATWAnomaly_r}
	\addplot [color=blue,style= thick,mark=triangle*,mark size=2.5,mark options={solid}] table[x index=0,y index=3] {\SWATNoiseTest}; \label{SWATWAnomaly_f1}
	\end{axis}
	
	\begin{axis}[  
	axis y line*=right,
	axis x line=none,
	legend style={nodes={scale=0.6}, fill opacity=0.6, text opacity=1},
	legend columns=2,
	legend pos = south west,
	height=4.5 cm,
	width=0.9\linewidth,
	font=\tiny,
	grid = major,
	ylabel={\# of detected attacks},
	ymin=23, ymax=33,
	ytick={23,24,25,26,27,28,29,30,31,32,33},
	yticklabels={23,24,25,26,27,28,29,30,31,32,33},
	ytick pos=right,
	enlargelimits=0.02,
	]
	\addlegendimage{/pgfplots/refstyle=SWATWAnomaly_p}\addlegendentry{Precision}
	\addlegendimage{/pgfplots/refstyle=SWATWAnomaly_r}\addlegendentry{Recall}
	\addlegendimage{/pgfplots/refstyle=SWATWAnomaly_f1}\addlegendentry{F1-score}
	\addplot [color=orange,dashed,style= thick,mark=star,mark size=3,mark options={solid}] table[x index=0,y index=4] {\SWATNoiseTest};
	\addlegendentry{\# attacks}
	\end{axis}
	\end{tikzpicture}
}

\newcommand{\chartWADINoise}{
	\begin{tikzpicture}
	\begin{axis}[  
	axis y line*=left,
	height=4.5 cm,
	width=0.9\linewidth,
	grid = both,
	xlabel={Noise $\sigma$},
	ylabel={Performance Metric},
	scaled y ticks=false,
	scaled x ticks=false,
	xmin=0,
	xmax=15,
	xtick={0,1,2,3,5,10,15},
	xticklabels={0,1,2,3,5,10,15},
	xtick pos=left,
	ymin=0.5, ymax=1,
	ytick={0.5,0.6,0.7,0.8,0.9,1.0},
	yticklabels={0.5,0.6,0.7,0.8,0.9,1.0},
	ytick pos=left,
	enlargelimits=0.02, 
	]
	\addplot [color=red,style= thick,mark=*,mark size=2,mark options={solid}] table[x index=0,y index=1] {\WADINoiseTest}; \label{WADIWAnomaly_p}
	\addplot [color=green!70!black,style= thick,mark=square*,mark size=2,mark options={solid}] table[x index=0,y index=2] {\WADINoiseTest}; \label{WADIWAnomaly_r}
	\addplot [color=blue,style= thick,mark=triangle*,mark size=2.5,mark options={solid}] table[x index=0,y index=3] {\WADINoiseTest}; \label{WADIWAnomaly_f1} \label{WADIWAnomaly_f1}
	\end{axis}
	
	\begin{axis}[  
	axis y line*=right,
	axis x line=none,
	legend style={nodes={scale=0.6}, fill opacity=0.6, text opacity=1},
	legend columns=2,
	legend pos = north east,
	height=4.5 cm,
	width=0.9\linewidth,
	grid = major,
	ylabel={\# of detected attacks},
	ymin=7, ymax=17,
	ytick={7,8,9,10,11,12,13,14,15, 16, 17},
	yticklabels={7,8,9,10,11,12,13,14,15, 16, 17},
	ytick pos=right,
	enlargelimits=0.02,
	]
	\addlegendimage{/pgfplots/refstyle=WADIWAnomaly_p}\addlegendentry{Precision}
	\addlegendimage{/pgfplots/refstyle=WADIWAnomaly_r}\addlegendentry{Recall}
	\addlegendimage{/pgfplots/refstyle=WADIWAnomaly_f1}\addlegendentry{F1-score}
	\addplot [color=orange,dashed,style= thick,mark=star,mark size=3,mark options={solid}] table[x index=0,y index=4] {\WADINoiseTest};
	\addlegendentry{\# attacks}
	\end{axis}
	\end{tikzpicture}
}

\newcolumntype{L}[1]{>{\raggedright\let\newline\\\arraybackslash\hspace{0pt}}m{#1}}
\newcolumntype{C}[1]{>{\centering\let\newline\\\arraybackslash\hspace{0pt}}m{#1}}
\newcolumntype{R}[1]{>{\raggedleft\let\newline\\\arraybackslash\hspace{0pt}}m{#1}}

In this section, we present a detailed evaluation of \ourtoolext\ obtained using the \ac{swat} and \ac{wadi} datasets presented in Section \ref{sec:datasets}. The evaluation comprises a comparison with state-of-the-art solutions in terms of detection accuracy, number of detected attacks and robustness to noisy data.

\subsection{Experimental Setup}\label{sec:Experimental_setup}

\ourtoolext\ has been implemented in PyTorch 1.0 \cite{pytorch-url} and validated using a Singularity container \cite{kurtzer2017singularity} running in a shared machine configured with 16 CPU cores, 64 GB virtual RAM and an NVIDIA 1080Ti GPU. The database of actuators normal states $A$ is implemented as a NumPy array populated using the training datasets. The maximum size of the array is less than 1MB for each of the two datasets.
Prior to our experiments, we also normalised the sensor readings between $0$ and $1$.

\subsection{Methodology}\label{sec:Methodology}

As per convention in the literature, we configure and evaluate \ourtoolext\ using the following metrics:
\begin{equation*}
\begin{gathered}
Pr=\frac{TP}{TP+FP}\quad Re=\frac{TP}{TP+FN}\quad F1=2\cdot\frac{Pr\cdot Re}{Pr + Re}
\end{gathered}
\end{equation*}

\noindent where \textit{Pr=Precision}, \textit{Re=Recall}, \textit{F1=F1 Score}, \textit{TP=True Positives}, \textit{FP=False Positives},  \textit{FN=False Negatives}. It is important to highlight that we adopt a point-based approach to compute these metrics. Thus, precision, recall and F1 score are computed using the labelled records (points) of the datasets, which report whether a record (a combination of actuators states and sensors readings recorded at a given time) is normal or anomalous. Like most of related works in the literature, we adopt this approach to tune the hyperparameters of \ac{wdnn} and \ac{ttnn} models, to assess the overall performance of \ourtoolext\ and for comparison with the state-of-the-art.

We used a grid search strategy to explore the set of hyper-parameters using F1 score as the performance metric. The final sets of hyper-parameters that maximise the F1 score on the two datasets are reported in Table \ref{tab:hyperparameters}.

\begin{table}[!ht]
	\centering
	\caption{Hyperparameters.}
	\label{tab:hyperparameters}
	\renewcommand{\arraystretch}{1.1}
	\begin{tabular}{L{3.4cm}C{2cm}C{2cm}}
	\hline
	\textbf{Hyperparameter} & \textbf{\ac{swat}} & \textbf{\ac{wadi}} \\ \hline
	Horizon ($H$) & 50 & 20   \\
	Input time window ($W_{in}$) & 60 & 50   \\
	Output time window ($W_{out}$) & 4 & 4   \\
	Anomaly time window ($W_{anom}$) & 30 & 30   \\
	\hline
\end{tabular}
\\
\vspace{1mm}
	\begin{tabular}{c}
		\ac{wdnn}
	\end{tabular}
\\
	\begin{tabular}{L{3.4cm}C{2cm}C{2cm}}
		\hline
		\textbf{Hyperparameter} & \textbf{\ac{swat}} & \textbf{\ac{wadi}} \\ \hline
		Learning rate ($\alpha$) & 0.01 & 0.001   \\
		Batch size ($s$) & 32 & 32   \\
		Tuning epochs ($EP_{tuning}$) & 100 & 100   \\
		DL1 Neurons & $W_{out}$ & $W_{out}$ \\
		DL2 Neurons & $3*W_{in}$ & $3*(m_{se}+m_{ac})$ \\ 
		DL3 Neurons & $W_{out}$ & $W_{out}$\\
		DL4 Neurons & 80 & 80\\
		DL5 Neurons & 2.25*$m^g_{se}$& 2.25*$m^g_{se}$\\
		DL6 Neurons & 1.5*$m^g_{se}$& 1.5*$m^g_{se}$ \\
		DL7 Neurons & $m^g_{se}$& $m^g_{se}$ \\
		CL1 Kernels, Kernel size & 64, 2  & 64, 5   \\
		MP1 Pooling size & (64, 2)  & (64, 2)   \\
		CL2 Kernels, Kernel size & 128, 2  & 128, 5   \\
		MP2 Pooling size & (128, 2)  & (128, 2)   \\
		\hline
	\end{tabular}
\\
\vspace{1mm}
\begin{tabular}{c}
	\ac{ttnn}
\end{tabular}
\\
\begin{tabular}{L{3.4cm}C{2cm}C{2cm}}
	\hline
	\textbf{Hyperparameter} & \textbf{\ac{swat}} & \textbf{\ac{wadi}} \\ \hline
	Learning rate ($\alpha$) &  0.01 & 0.01   \\
	Batch size ($s$) & 32 & 32   \\
	Tuning epochs ($EP_{tuning}$) & 1 & 1   \\
	Median kernel size & 59 & 59   \\
	CL1 Kernels, Kernel size & 2, 2  & 2, 2   \\
	MP1 Pooling size & (2, 2)  & (2, 2)   \\
	CL2 Kernels, Kernel size & $W_{out}$, 2  & $W_{out}$, 2   \\
	MP2 Pooling size & ($W_{out}$, 2)  & ($W_{out}$, 2)   \\
	DL Neurons & 1 & 1\\
	\hline
\end{tabular}
\end{table}

These hyper-parameters are kept constant throughout our experiments presented below in this section, except when we study the sensitivity of \ourtoolext\ to a specific hyper-parameter. 

In our experiments, we measure the sensitivity of \ourtoolext\ to noise. It is worth recalling that industrial control systems operate in hostile environments \cite{ODVA2016EtherNet}, where the communication channels are often subject to interference (e.g., in the case of employing wireless communication devices \cite{Galloway2013}).
For evaluation purposes we add Gaussian noise to sensor readings of the \ac{swat} dataset with mean $\mu=0$ and standard deviation $\sigma\in\{ 1, 2, 3, 5, 10, 15 \}$. The noise distribution and the values of $\mu$ and $\sigma$ have been selected based on similar assumptions made in other works  regarding the noise in communication channels of networked control systems~\cite{goodwin2008architectures, zhan2012performance}.

In this research, we do not evaluate the robustness of \ourtoolext\ against adversarial noise, i.e., the noise added to the communication channels by an attacker intending to perturb the detection accuracy. Indeed, the defence against adversarial attacks in \acp{ics} is a complex problem, and its discussion is outside the scope of this work.

The validation presented below is divided into three different experiments. In Experiment 1 and 2, we compare the performance of \ourtoolext\ with relevant works in the state-of-the-art in terms of precision, recall, F1 score, and the number of attacks correctly detected. Experiment 1 evaluates \ourtoolext\ using the \ac{swat} dataset, while experiment 2 uses the \ac{wadi} dataset. In Experiment 3, we evaluate the robustness of \ourtoolext\ to additive noise applied on the \ac{swat} and \ac{wadi} test sets. We also compare the results with the frameworks proposed in \cite{Kravchik2018, abdelaty2020aads} in case of the \ac{swat} dataset.

\subsection{Experiment 1: Detection accuracy in the \ac{swat}}\label{sec:Exp_1}
As shown in Table \ref{tab:swat_results_comparison}, \ourtoolext\ outperforms existing state-of-the-art detection on the \ac{swat} test set with $88.9\%$ F1 score, and correctly recognises 33 out of 36 attacks in the test set. It is worth noting that \ourtoolext\ also detects a higher rate of attacks during their execution (97\% against 93\% of \cite{Kravchik2018}), hence allowing the operator to activate the adequate countermeasures more promptly.

\newcommand{\SwatHIntB}{645}
\newcommand{\SwatHIntS}{192}
\begin{table}[!ht]
	\centering
	\caption{State-of-the-art comparison (\ac{swat} dataset). In parenthesis, the number of attacks detected after their end.}
	\label{tab:swat_results_comparison}
	\renewcommand{\arraystretch}{1.1}
	\begin{tabular}{l@{\hskip 1cm}c@{\hskip 3mm}c@{\hskip 3mm}c@{\hskip 3mm}c}
		\hline
		\textbf{Architecture} & \textbf{Precision} & \textbf{Recall} & \textbf{F1} & \begin{tabular}{@{}c@{}}\textbf{Detected}\\\textbf{attacks}\end{tabular} \\ \hline
		\textbf{\ourtoolext} & 0.9185 & 0.8616 & 0.8892 & 33(1)  \\
		\textbf{\ourtool} \cite{abdelaty2020aads} & 0.866 & 0.861 & 0.863 & 33(1)  \\
		MLP \cite{Shalyga2018}& 0.967& 0.696 &0.812&  25 \\
		\ac{cnn} \cite{Kravchik2018} &0.867 & 0.854 & 0.860& 31(2) \\
		TABOR \cite{lin2018tabor} &0.861 & 0.788 & 0.823 & 24 \\
		Windowed-PCA \cite{kravchik2019efficient} &0.92 & 0.841 & 0.879 & - \\
		Online-ADS \cite{FARSI2019online} &0.8638 & 0.9064 & 0.8846 & - \\
		\hline
	\end{tabular}
\end{table}

\textbf{Few-time-steps algorithm.} We also evaluated the contribution of the few-time-steps algorithm to the detection accuracy of \ourtoolext. To this aim, we repeated the experiment disabling the algorithm. As the \ac{swat} test set contains normal records that are not present in the training set, the model of normal behaviour built with the training set leads to several false positives and, more precisely, to a low F1-score measure of $77.7\%$ and $1251$ false alarms (compared to the $645$ experienced with the algorithm enabled). 

The update process is fast. Indeed, with our setup, the execution of one cycle of the few-time-steps learning algorithm, including the $EP_{tuning}=100$ epochs, with a batch of 32 samples takes 400 ms on average. 
In an online system, the algorithm can be triggered by the operator upon identifying a false alarm (e.g., a known unstable sensor or actuator). It is important to stress that this is the only requirement for the operator, unlike other solutions where an in-depth knowledge of the underlying algorithms is necessary to update thresholds or other parameters.

\begin{figure}[h!]
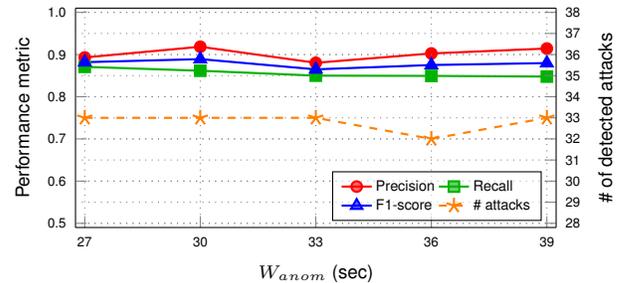

	\centering
	\chartSWATWAnomalyText%
	\caption{Sensitivity of \ac{wdnn} to $W_{anom}$ on \ac{swat} dataset.}
	\label{fig:swat_w_anomaly}
\end{figure}

\textbf{Hyper-parameters.} The sensitivity of \ourtoolext\ to the hyper-parameter $W_{anom}$ is presented in Figure \ref{fig:swat_w_anomaly}. We remind that $W_{anom}$ is defined in Equation \ref{eq:anomaly_label} as the number of consecutive times the prediction error \ac{mse} is higher of the anomaly threshold on the same output section of \ac{wdnn}.   

Although the number of detected attacks is quite stable to 33 (except for $W_{anom}$=36), the highest F1 score (0.8892) is measured at $W_{anom}$=30.

\textbf{False alarms.} \ourtoolext\ relies on the feedback from the technician to fine-tune the output sections with new information of the normal behaviour. Of course, the rate of technician's interventions is a relevant metric to consider, as too frequent false positives can undermine the usability of the system. With the chosen settings, we measured 645 false alarms in total on the test set of the \ac{swat} dataset, equivalent to 6.6 human interventions/hour on average. Although this seems a manageable rate of interventions, in more complex industrial environments compared to the \ac{swat} testbed.    

One way to reduce the rate of intervention is to allow a \textit{grace time} $W_{grace}$, through which alarms are reported to the technician only if they last for at least $W_{grace}$ seconds. Of course, this practice may hide short true positives, hence preventing the detection of a certain number of anomalies. 

\begin{figure}[t!]
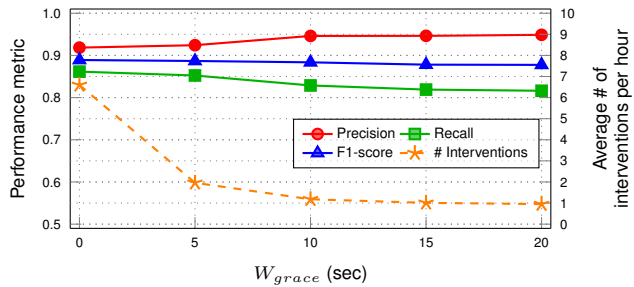

	\centering
	\chartSWATHIntWSilentText%
	\caption{Sensitivity of \ac{wdnn} to $W_{grace}$ on \ac{swat} dataset.}
	\label{fig:swat_w_grace}
\end{figure}

We tested the sensitivity of \ourtoolext\ to the grace time by varying  $W_{grace}$ from 1 to 20 seconds. In Figure \ref{fig:swat_w_grace} we can observe a minimal decrease of the F1 score (0.877 with $W_{grace}$=20) and a consistent reduction of the technician intervention rates (from 6.6/hour with no grace time, to 1.96/hour with $W_{grace}$=5, to 0.95/hour with $W_{grace}$=20). Moreover, we did not experience any variations in the number of detected anomalies with $W_{grace}$=5 (still 33), while with $W_{grace}$=20 this number decreases to 28, as somehow expected.

\subsection{Experiment 2: Detection accuracy in the \ac{wadi}}
\label{sec:Exp_2}
\ourtoolext\ matches the state of the art results in terms of number of attacks detected, 14 out of 15 attacks in the \ac{wadi} dataset, while it outperforms the other solutions in terms of F1 score, as summarised in Table \ref{tab:results-comparison-wadi}.

\newcommand{\WadiHIntB}{226}
\newcommand{\WadiHIntS}{110}
\begin{table}[!ht]
	\centering
	\caption{State-of-the-art comparison (\ac{wadi} dataset).}
	\label{tab:results-comparison-wadi}
	\renewcommand{\arraystretch}{1.1}
	\begin{tabular}{l@{\hskip 1cm}c@{\hskip 3mm}c@{\hskip 3mm}c@{\hskip 3mm}c}
		\hline
		\textbf{Architecture} & \textbf{Precision} & \textbf{Recall} & \textbf{F1} & \begin{tabular}{@{}c@{}}\textbf{Detected}\\\textbf{attacks}\end{tabular} \\ \hline
		\textbf{\ourtoolext} & 0.9083 & 0.7205 & 0.8036 & 14  \\
		MAD-GAN \cite{li2019madgan} &0.4144 & 0.3392 & 0.37 & - \\
		1D CNN \cite{kravchik2019efficient} &0.697 & 0.731 & 0.714 & 14 \\
		AE \cite{kravchik2019efficient} &0.834 & 0.681 & 0.750 & 14 \\
	\hline
	\end{tabular}
\end{table}

\textbf{Few-time-steps algorithm.} Given the setup described in section \ref{sec:Experimental_setup}, \ourtoolext\ takes 800 ms to fine-tune an output section with 100 epochs and a batch of 32 samples. Like in the case of the \ac{swat} dataset, we demonstrate how the few-time-steps algorithm contributes to keeping weights and biases of \ac{wdnn} up-to-date with the changes of the normal behaviour. Indeed, when disabling the updating mechanism, the F1-score drops to 0.5621, while the number of false alarms increases to from \WadiHIntB\ to 684.

\begin{figure}[h!]
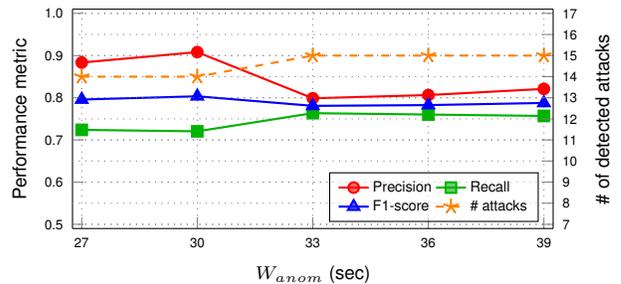

	\centering
	\chartWADIWAnomalyText%
	\caption{Sensitivity of \ac{wdnn} to $W_{anom}$ on \ac{wadi} dataset.}
	\label{fig:wadi_w_anomaly}
\end{figure}

\textbf{Hyper-parameters.} Increasing the anomaly window $W_{anom}$ determines an increase on the number of detected attacks, as shown in Figure \ref{fig:wadi_w_anomaly}. This is mainly due to the condition $\ac{mse}_{g,i} > T_g$ in Equation \ref{eq:anomaly_label}, which yields to the detection of short attacks when $W_{anom}$ is large enough. However, for the state-of-the-art comparison reported in Table \ref{tab:results-comparison-wadi}, we set $W_{anom}$=30, as this value maximises the F1 score.

\begin{figure}[h!]
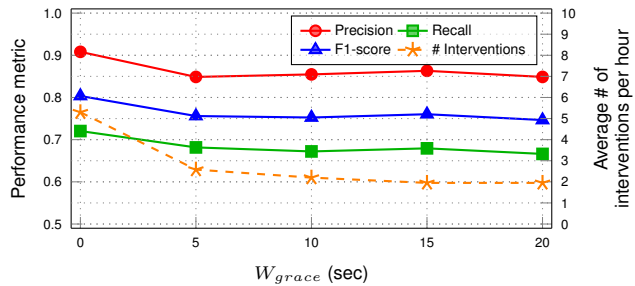

	\centering
	\chartWADIHIntWSilentText%
	\caption{Sensitivity of \ac{wdnn} to $W_{grace}$ on \ac{wadi} dataset.}
	\label{fig:wadi_w_grace}
\end{figure}

\textbf{False alarms.} With the settings listed in Table \ref{tab:hyperparameters}, \ourtoolext\ produces \WadiHIntB\ false positives on the \ac{wadi} test set, equivalent to 5.3 human interventions per hour on average. As for the \ac{swat} dataset, we measured the sensitivity of \ourtoolext\ to the \textit{grace time} $W_{grace}$. The results, reported in Figure \ref{fig:wadi_w_grace}, show that increasing $W_{grace}$ reduces the number of intervention to 1.95 per hour on average at $W_{grace}$=20, although negatively impacting on precision, recall and F1 score even at $W_{grace}$=5.

\subsection{Experiment 3: robustness to additive noise}\label{sec:Exp_3}

In this experiment, we measure the robustness of \ourtoolext\ to noise added to the sensors readings. In our experiments, we add various levels synthetic noise to the sensor readings of both \ac{swat} and \ac{wadi} datasets. As anticipated in Section \ref{sec:Methodology}, we use white Gaussian noise with mean $\mu=0$ and increasing standard deviation $\sigma\in\{ 1, 2, 3, 5, 10, 15 \}$.  We then observe the behaviour of \ourtoolext\ with respect to the number of detected attacks and the detection accuracy measured with the F1 score metric. We then extend the experiment 2 in \cite{abdelaty2020aads}, where \ourtool\ and the \ac{cnn} proposed in \cite{Kravchik2018} are compared, by adding the performance of \ourtoolext\ on the \ac{swat} dataset.

\begin{figure*}[t!]
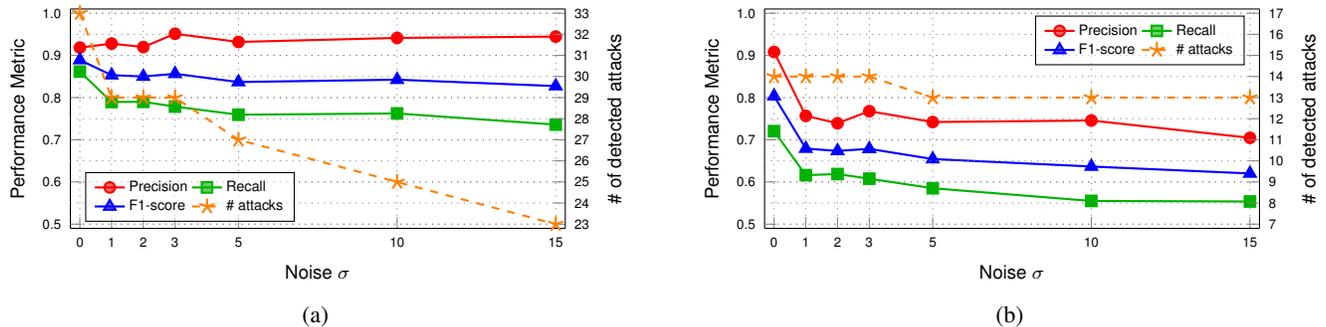

	\centering
	\begin{subfigure}[t]{0.5\textwidth}
		\centering
		\chartSWATNoise%
		\caption{}
		\label{fig:swat_noise}
	\end{subfigure}%
	~ 
	\begin{subfigure}[t]{0.5\textwidth}
		\centering
		\chartWADINoise%
		\caption{}
		\label{fig:wadi_noise}
	\end{subfigure}
	\caption{Performance of \ourtoolext\ when adding Gaussian noise to the \ac{swat} (a) and \ac{wadi} (b) test sets.}
	\label{fig:noise}	
	\vspace{-2mm}
\end{figure*}

Figures \ref{fig:swat_noise} and \ref{fig:wadi_noise} report on the performance of \ourtoolext\ as a function of the Gaussian noise level. 
Although the trend of the point-based F1 score is similar on both datasets, we can observe a higher impact of the synthetic noise on the detection of attacks in the \ac{swat} dataset. We recall that an attack is determined when the conditions in Equation \ref{eq:anomaly_label} are met. In the case of sensors, the \ac{mse} must be above the threshold for $W_{anom}$ consecutive samples. However, the increasing number of point-based false negatives introduced with the noise, as shown in the two figures by the trend of the recall measure, reduces the chances of finding $W_{anom}$ consecutive point-based positive samples. Of course, this affects more the detection of short attacks, which are more frequent in the \ac{swat} dataset.

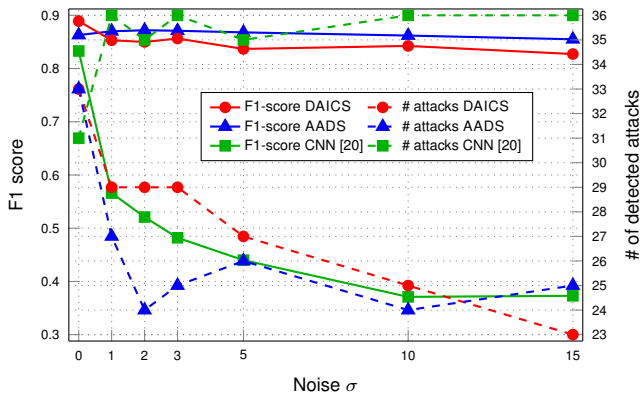
\begin{figure}[h!]
	\centering
	\begin{tikzpicture}
	\begin{axis}[  
	axis y line*=left,
	height=6 cm,
	width=0.95\linewidth,
	grid = both,
	xlabel={Noise $\sigma$},
	ylabel={F1 score},
	scaled y ticks=false,
	scaled x ticks=false,
	xmin=0,
	xmax=15,
	xtick={0,1,2,3,5,10,15},
	xticklabels={0,1,2,3,5,10,15},
	xtick pos=left,
	ymin=0.3, ymax=0.9,
	ytick={0.3,0.4,0.5,0.6,0.7,0.8,0.9},
	yticklabels={0.3,0.4,0.5,0.6,0.7,0.8,0.9},
	ytick pos=left,
	enlargelimits=0.02, 
	]
	\addplot [color=blue,style=thick,mark=triangle*,mark size=2.5, mark options={solid}] table[x index=0,y index=1] {\NoiseTest}; \label{ourF1}
	\addplot [color=green!70!black,style=thick,mark=square*,mark size=1.8, mark options={solid}] table[x index=0,y index=2] {\NoiseTest}; \label{cnnF1}
	\addplot [color=red,style= thick,mark=*,mark size=1.8, mark options={solid}] table[x index=0,y index=5] {\SWATNoiseTestMedfilterFS}; \label{our_extF1}
	\end{axis}
	
	\begin{axis}[  
	axis y line*=right,
	axis x line=none,
	legend style={nodes={scale=0.6}, at={(0.58,0.75))},anchor=north},
	legend columns=2,
	height=6 cm,
	width=0.95\linewidth,
	grid = major,
	ylabel={\# of detected attacks},
	ymin=23, ymax=36,
	ytick={23,24,25,26,27,28,29,30,31,32,33,34,35,36},
	yticklabels={23,24,25,26,27,28,29,30,31,32,33,34,35,36},
	ytick pos=right,
	enlargelimits=0.02,
	]
	\addlegendimage{/pgfplots/refstyle=our_extF1}\addlegendentry{F1-score \ourtoolext}
	\addplot [color=red,dashed,style=thick,mark=*,mark size=1.8, mark options={solid}] table[x index=0,y index=5] {\SWATNoiseTestMedfilterAtt};
	\addlegendentry{\# attacks \ourtoolext}
	\addlegendimage{/pgfplots/refstyle=ourF1}\addlegendentry{F1-score \ourtool}
	\addplot [color=blue,dashed,style=thick,mark=triangle*,mark size=2.5, mark options={solid}] table[x index=0,y index=1] {\AttacksTest};
	\addlegendentry{\# attacks \ourtool}
	\addlegendimage{/pgfplots/refstyle=cnnF1}\addlegendentry{F1-score CNN\cite{Kravchik2018}}
	\addplot [color=green!70!black,dashed,style=thick,mark=square*,mark size=1.8, mark options={solid}] table[x index=0,y index=2] {\AttacksTest};
	\addlegendentry{\# attacks CNN\cite{Kravchik2018}}
	\end{axis}
	\end{tikzpicture}
	\caption{Performance when adding Gaussian noise to the \ac{swat} test set.}
	\label{fig:experiment_3}
\end{figure}

Figure \ref{fig:experiment_3} shows the values of F1-score and number of detected attacks as functions of $\sigma$. We can notice that \ourtoolext\ improves \ourtool\ in the number of detected attacks at any level of noise, except for $\sigma$=15 while almost matching the F1 score (we measured a slight decrease of 0.015 on average for $\sigma>$ 0).  This demonstrates that the new mechanism is solid also in noisy conditions. On the contrary,  the F1 score of the state of the art \ac{cnn} drops drastically (green solid curve). This is mainly due to the statistical approach and to the static threshold employed to detect the anomalies, empirically selected as the value $T\in[1.8,3]$ that maximizes the F1-score. The low values of the F1-Score for $\sigma>0$ are due to a low precision measure (around 0.31 on average), meaning that the \ac{cnn} classifies most of the records as anomalies, mostly false positives. This is the reason why the 36 attacks in the \ac{swat} test set are almost all correctly classified (green dashed line). However, as everything looks like an anomaly, in a real-world deployment, the output of the \ac{cnn} would be unusable. \\

In conclusion, \ourtoolext\ overcomes state-of-the-art solutions on both \ac{swat} and \ac{wadi} datasets with respect to point-based F1 score and number of detected attacks. In addition, \ourtoolext\ is more robust to the additive noise, hence potentially more reliable in real-world industrial scenarios where the electromagnetic noise and the natural degradation of the devices might result in noisy data.

\section{Conclusion}\label{Conclusion}

In this paper, we have presented \ourtoolext, a framework for anomaly detection in \acfp{ics} grounded on the one-class classification paradigm. \ourtoolext\ has been designed to mitigate two major problems affecting similar solutions proposed in the state-of-the-art literature. The first, called \textit{domain shift}, can be observed when changes in the normal behaviour of the \ac{ics} are not correctly handled by the detection system. The result is often an increase in the number of false alarms. The second problem is caused by the presence of noisy data, either due to interference on the communication channel within the \ac{ics} or to the ageing of devices, which prevents the detection system from correctly segregating the anomalies from normal operations.

We have tackled the aforementioned problems by introducing a fast and automated mechanism for updating the \ac{ics} model based on the detected false alarms caused by changes in the normal behaviour. Such a mechanism, based on the so-called \textit{few-time-steps algorithm}, has been designed to be usable in production environments. Indeed, the only assumption is that the technician can identify the false alarms caused by maintenance operations (hardware/software updates) or known misbehaving devices, and to signal them to \ourtoolext. This  sets \ourtoolext\ apart from similar solutions, where an in-depth knowledge of the underlying algorithms is necessary to update thresholds or other parameters. The combination of the few-time-steps algorithm with a dynamic threshold mechanism, also proposed in this work, allows \ourtoolext\ to overcome state-of-the-art solutions in terms of number of detected attacks and resilience to noisy data samples. 



\section*{Acknowledgment}
The authors would like to thank the Center for Research in Cyber Security at the Singapore University of Technology and Design for providing the \ac{swat} and \ac{wadi} datasets.

\bibliography{bibliography.bib}
\bibliographystyle{IEEEtran}

\end{document}